\documentclass[fleqn,usenatbib]{mnras}

\usepackage{hyperref}

\usepackage[T1]{fontenc}
\usepackage{ae,aecompl}

\usepackage{graphicx}	% Including figure files
\usepackage{amsmath}	% Advanced maths commands, needed for astro-ph compilation
\usepackage{amssymb}	% Extra maths symbols, needed for astro-ph compilation
\usepackage{multirow}
\usepackage{subfig,float}

\usepackage{colortbl}

\newcommand{\kms}{km\,s$^{-1}$}

\newcommand{\vsini}{$v\sin{i}$}

\title[Magnetic geometry and latitudinal differential rotation of Alhena]{Magnetic geometry and surface differential rotation of the bright Am star Alhena A\thanks{Based on observations obtained at the T\'elescope Bernard Lyot (TBL) at Observatoire du Pic du Midi, CNRS/INSU and Universit\'e de Toulouse, France}}

\author[A. Blaz\`ere et al.]{
Aurore Blaz\`ere$^{1,2}$\thanks{E-mail: ablazere@ulg.ac.be}, 
Pascal Petit$^{3}$,
Coralie Neiner$^{2}$,
Colin Folsom$^{3}$,
Oleg Kochukhov$^{4}$,
\newauthor St\'ephane Mathis$^{5,2}$,
Morgan Deal$^{2}$,
John Landstreet$^{6}$
\\
$^{1}$Institut d'Astrophysique et de G\'eophysique, Universit\'e de Li\`ege, Quartier Agora (B5c), All\'ee du 6 ao\^ut 19c, 4000 Sart Tilman, Li\`ege, Belgium\\
$^{2}$LESIA, Observatoire de Paris, PSL University, CNRS, Sorbonne Universit\'e, Univ. Paris 
Diderot, Sorbonne Paris Cit\'e,\\ 5 place Jules Janssen, 92195 Meudon, France\\
$^{3}$IRAP (Institut de Recherche en Astrophysique et Plan\'etologie), Universit\'e de Toulouse, CNRS, CNES, UPS, Toulouse, France\\
$^{4}$Department of Physics and Astronomy, Uppsala University, Box 516, SE-751 20 Uppsala, Sweden\\
$^{5}$Laboratoire AIM Paris-Saclay, CEA/DRF - CNRS - Universit\'e Paris Diderot, IRFU/DAp Centre de Saclay, F-91191 Gif-sur-Yvette, France\\
$^{6}$Dept. of Physics \& Astronomy, University of Western Ontario, London, Ontario, Canada
}

\date{Accepted XXX. Received YYY; in original form ZZZ}
\pubyear{2017}

\begin{document}
\label{firstpage}
\pagerange{\pageref{firstpage}--\pageref{lastpage}}
\maketitle

% Abstract of the paper
\begin{abstract}

Alhena A ($\gamma$ Gem A) is a bright Am star, with the strongest disc-integrated magnetic field strength reported so far for an Am star. Its spectrum exhibits standard circularly polarized Zeeman signatures, contrary to all previously studied Am stars that display abnormal signatures dominated by a single-signed lobe. 

We present here the result of follow-up observations of Alhena, using very high signal-to-noise spectropolarimetric data obtained over 25 observing nights with NARVAL at T\'elescope Bernard Lyot, in the frame of the BRITE (BRIght Target Explorer) spectropolarimetric survey.

We confirm that Alhena A is magnetic and we determine its surface magnetic properties using different methods. Inclined dipole models are used to reproduce the longitudinal field measurements, as well as the Stokes V line profiles themselves. In both cases, the model is consistent with a polar field strength of $\sim$ 30 G. This is confirmed by a Zeeman-Doppler Imaging (ZDI) model, which also unveils smaller scale magnetic structures. A rotational period of 8.975 days was identified using intensity line profile variations. The ZDI inversion suggests that the surface magnetic field is sheared by differential rotation, with a difference in rotation rate between high and low latitudes at about 15\% of the solar value. This result challenges theories of the development of surface differential rotation in intermediate mass main sequence stars.

\end{abstract}


%%%%%%%%%%%%%%%%%%%%%%%%%%%%%%%%%%%%%%%%%%%%%%%%%%

\section{Introduction}
\label{intro}

Stellar magnetic fields have a strong influence on stellar evolution, being involved in a number of physical processes operating within and in the immediate vicinity of stars, such as the accretion, diffusion, and winds. Until recently, among intermediate-mass stars, only the chemically peculiar Ap/Bp stars were known to be magnetic (e.g. \citealt{auriere07,auriere10}). These stars feature a simple field geometry (mostly dominated by a dipole), with strong magnetic field strengths (B$_d$ $\geq$ 300 G). Their magnetic geometries are found to be stable on time scales of several decades \citep{silvester14}. The  scenario commonly accepted to explain the origin of this magnetism is the fossil field theory. In this framework, magnetic fields observed at the surface of hot stars are proposed to be a remnant of magnetic fields amplified during an earlier evolutionary phase (\citealt{mestel99}, \citealt{neiner2015}, or \citealt{braithwaite15} for a review and the description of alternate theories). 

This understanding of magnetic fields in tepid stars was enriched by the discovery of an ultra-weak magnetic field (longitudinal magnetic field below 1 gauss) at the surface of the fast rotating, $\lambda$~Bo\"otis star Vega \citep{lignieres09,petit10}. This discovery raised the question of the possible ubiquity of weak fields in intermediate-mass stars that do not host strong magnetic fields. It also lead to a new vision of intermediate-mass star magnetism in which two types of magnetism may coexist, with strong magnetic fields affecting Ap/Bp stars, and ultra-weak fields like the one of Vega, separated by a magnetic desert of about two orders of magnitude in field strength. To explain this dichotomy, two main scenarios have been developed. The first scenario was based on the stability of magnetic configurations  at large scale in a differentially rotating star. Above a critical field strength, magnetic fields can remain stable on long timescales, while  below this limit magnetic fields would likely be destroyed by instability, producing ultra-weak magnetic fields \citep{auriere07}. The other scenario, called failed fossil fields, proposes that the ultra-weak magnetic fields are the result of a field evolving dynamically towards equilibrium with a timescale to reach it longer than the lifetime of the star \citep{brait13}. These scenarios required new observational constraints. Each discovery and study of an ultra-weak field could bring new clues about the dichotomy between strong and weak fields and improve the scenario to explain it.
                                                          
In addition, ultra-weak magnetic field signatures have been detected in three Am stars: Sirius\,A \citep{petit11}, $\beta$\,UMa, and $\theta$\,Leo \citep{blazere16a}, thanks to very high signal-to-noise spectropolarimetric observations. These three stars exhibit abnormal signatures in circular polarization, with a prominent positive lobe dominating over the negative one. Such signatures, although not expected in the standard Zeeman effect theory, were demonstrated to be magnetic \citep{blazere16a}. A preliminary explanation for these peculiar signatures is the combination of a strong vertical gradient of velocity with a gradient of magnetic field in the superficial convective layer of the star. 

A first normal magnetic signature with a positive and negative lobe, like the one of Vega,  was discovered on the Am star Alhena (\citealt{blazere16b}, hereafter B16). The difference between the field of Alhena and the other Am stars is puzzling. In particular, the stellar parameters of Alhena are very similar to the ones of $\theta$ Leo, which exhibits peculiar signatures. One explanation proposed by B16 is based on the low micro-turbulence of Alhena compared to other Am stars.  A low micro-turbulence can be interpreted as due to the absence of a superficial convective shell, which is the most likely physical ingredient needed to produce very asymmetric Stokes V signatures \cite{landstreet09,lopez02}. Alhena is thus a very interesting star, which might provide the clue to understanding the peculiar shapes of the magnetic signatures of the other Am stars. In this paper, we further investigate the magnetic properties of Alhena, by presenting the result of a spectropolarimetric follow up of Alhena.

Alhena is a well known bright spectroscopic binary. The primary was classified as a subgiant A0IV star \citep{gray14} with a T$_{\rm eff}$= 9150$\pm$310 K  and $\log$ g=3.60$\pm$0.20 \citep{adelman15} and the secondary was classified as a cool G star \citep{thalmann14}. The orbital elements of the binary have been determined thanks to AO observations and speckle interferometry \citep{drummond14}. The orbit of Alhena is very eccentric ($e$=0.89) with a period of 12.63 years. \cite{fekel93} estimated that the mass of the primary is 2.8 M$_{\odot}$ and the one of the secondary is 1.07 M$_{\odot}$ . The primary, Alhena\,A, is classified as a normal star, with some over abundances heading towards an Am star, by \cite{adelman15}. This suggests that Alhena\,A is between a normal and a weakly Am star.

Based on new spectropolarimetric observations of Alhena presented in Sect.~\ref{sect_obs}, we here seek to confirm that the primary (the Am star) Alhena A is a magnetic star. We first  present information on the binarity of Alhena (Sect.~\ref{binary}) and our magnetic measurements (Sect.~\ref{sect_mag}). We then determine the strength of the magnetic field of Alhena A (Sect.~\ref{sect_field}) and the properties of the magnetic field using several methods. Finally, we discuss our results in Sect.~\ref{discus} and draw conclusions in Sect.~\ref{conclu}.

\section{Spectropolarimetric observations}
\label{sect_obs}

Observations were obtained with the NARVAL spectropolarimeter (\citealt{auriere03}, \citealt{silvester12}) as part of the BRITEPol spectropolarimetric project \citep{neiner16}. This survey was performed as a complement to the BRITE-Constellation project, which explores asteroseismology of stars with V$\leq$4 thanks to a constellation of nano-satellites \citep{weiss14}. The aim of the BRITEPol ground-based program is to obtain spectropolarimetric observations of BRITE targets to offer systematic magnetic field measurements and high resolution, high signal-to-noise ratio spectra of stars for which seismic constraints will be obtained through BRITE data. Unfortunaly, Alhena was not observed with the BRITE constellation so far.

NARVAL is a high-resolution spectropolarimeter,installed at the 2-meter Bernard Lyot Telescope (TBL) at the Pic du Midi Observatory. This spectropolarimeter was optimized to detect stellar magnetic fields thanks to the polarization signatures they generate in photospheric spectral lines. It covers, in one single exposure, a wide wavelength domain from about 375 to 1050 nm, on 40 echelle orders with a spectral resolution of $\sim$65000.

Alhena was observed in polarimetric mode to measure the circular polarization (Stokes V). Each observation was divided in 4 sub-exposures, taken in different configurations of the polarimeter retarders. In addition to the intensity (Stokes I) spectra, the Stokes V spectra were obtained by combining constructively the 4 sub-exposures.  To check for spurious detection due to e.g., instrumental effects, variable observing conditions, or non-magnetic physical effects such as pulsations, a null polarization (N) spectrum was obtained by destructively combining the sub-exposures. Alhena was observed once on October 27, 2014, 19 times between September 2015 and April 2016 and 5 times in April/May 2017. During the last 9 nights, 3 successive observations were obtained each night. In total, we obtained 43 observations  taken on 25 nights over several years. The journal of observations is provided in Table~\ref{journal_alhena}.

\begin{table*}
\caption{Journal of observations of Alhena indicating the date of observation, Heliocentric Julian Date at the middle of the observations (mid-HJD - 2450000), radial velocity in km\,s$^{-1}$, the number of sequences and exposure time in seconds, the mean S/N of the intensity spectrum at $\sim$500 nm, the orbital phase and the rotational phase.} 
\centering
\begin{tabular}{l c c c c c c c}
\hline
\# & date & mid-HJD & RV & $T_{\rm exp}$ & S/N & Phase$_{orb}$ & Phase$_{rot}$\\
\hline
\hline
 1 & 27 Oct. 14 & 6958.6531 & $-15.58 \pm 0.14$ &4 $\times$ 25 & 986&	 0.7815		    & 0.212\\
 2 & 18 Sep. 15 & 7284.6954 & $-16.03 \pm 0.24$ &4 $\times$ 35 & 1016&   0.8790		    & 0.499\\
 3 & 19 Sep. 15 & 7285.6944 & $-15.96 \pm 0.25$ &4 $\times$ 35 & 1093&  0.8793		   & 0.610\\
 4 & 08 Oct. 15 & 7304.7204 & $-16.13 \pm 0.26$ &4 $\times$ 35 & 1152&  0.8834		    & 0.728\\
 5 & 09 Oct. 15 & 7305.7266 & $-16.19 \pm 0.16$ &4 $\times$ 35 & 1194&  0.8836		    & 0.839\\
 6 & 10 Oct. 15 & 7306.7104 & $-16.22 \pm 0.18$ &4 $\times$ 35 & 961&	0.8838		    & 0.949\\
 7 & 14 Oct. 15 & 7310.5895 & $-16.31 \pm 0.18$ &4 $\times$ 35 & 938&	 0.8847		   & 0.381\\
 8 & 20 Oct. 15 & 7316.6682 & $-16.36 \pm 0.14$ &4 $\times$ 35 & 832&	 0.8860		   & 0.058\\
 9 & 30 Oct. 15 & 7326.7289 & $-16.34 \pm 0.17$ &4 $\times$ 35 & 1157&   0.8882		    & 0.177\\
10 & 31 Oct. 15 & 7327.7354 & $-16.37 \pm 0.15$ &4 $\times$ 35 & 1149& 0.8884		    & 0.289\\
11 & 09 Nov. 15 & 7336.7300 & $-16.39 \pm 0.20$ &4 $\times$ 35 & 935 &  0.8903		    & 0.290\\
12 & 16 Nov. 15 & 7343.6352 & $-16.49 \pm 0.12$ &4 $\times$ 35 & 917&	0.8918		    & 0.059\\
13 & 01 Dec. 15 & 7358.6118 & $-16.24 \pm 0.25$ &4 $\times$ 35 & 951&	 0.8951		    & 0.726\\
14 & 06 Dec. 15 & 7363.6642 & $-16.45 \pm 0.18$ &4 $\times$ 35 & 1320&   0.8962		    & 0.288\\
15 & 11 Dec. 15 & 7368.6264 & $-16.31 \pm 0.15$ &4 $\times$ 35 & 1170&  0.8972		    & 0.841\\
16 & 17 Dec. 15 & 7374.6085 & $-16.13 \pm 0.22$ &4 $\times$ 35 & 1057& 0.8985		    & 0.506\\
17 & 20 Jan. 16 & 7408.6078 & $-16.54 \pm 0.19 $ &3 $\times$ 4 $\times$ 42 & 2246& 0.9059    & 0.290\\
18 & 20 Feb. 16 & 7439.4452 & $-16.42 \pm 0.23$ &3 $\times$ 4 $\times$ 42 & 1323& 0.9126     & 0.722\\
19 & 20 Mar. 16 & 7460.4035 & $-16.74 \pm 0.15$ &3 $\times$ 4 $\times$ 42 & 2307& 0.9171     & 0.054\\
20 & 06 Apr. 16 & 7485.3300 & $-16.57 \pm 0.16$ &3 $\times$ 4 $\times$ 42 & 2173& 0.9225     & 0.829\\
21 & 20 Apr. 17 & 7864.3198 & $6.99 \pm 0.13 $ &3 $\times$ 4 $\times$ 42 & 1028&  0.0046     & 0.010\\
22 & 21 Apr. 17 & 7865.3263 & $6.91 \pm 0.17  $ &3 $\times$ 4 $\times$ 42 & 1346& 0.0049     & 0.122\\
23 & 22 Apr. 17 & 7866.3196 & $6.82 \pm 0.18 $ &3 $\times$ 4 $\times$ 42 & 1203&  0.0051     & 0.232\\
24 & 03 May 17  & 7877.3349 & $5.92 \pm 0.21 $ &3 $\times$ 4 $\times$ 42 & 1141&  0.0084     & 0.458\\ 
25 & 07 May 17  & 7881.3275 & $5.08  \pm 0.19$ &3 $\times$ 4 $\times$ 42 & 1347&  0.0096    & 0.902\\
\hline
\end{tabular}
\label{journal_alhena}
\end{table*}

The data reduction was performed at the telescope using the Libre-Esprit reduction package
\citep{donati97}. Each of the 40 echelle orders of each of
the 43 spectra were normalized using the \texttt{continuum} task of IRAF\footnote{IRAF is distributed by
the National Optical Astronomy Observatories, which are operated by the
Association of Universities for Research in Astronomy, Inc., under cooperative
agreement with the National Science Foundation.}.

\section{Binarity}
\label{binary}

\begin{figure}
\centering
\includegraphics[width=8cm]{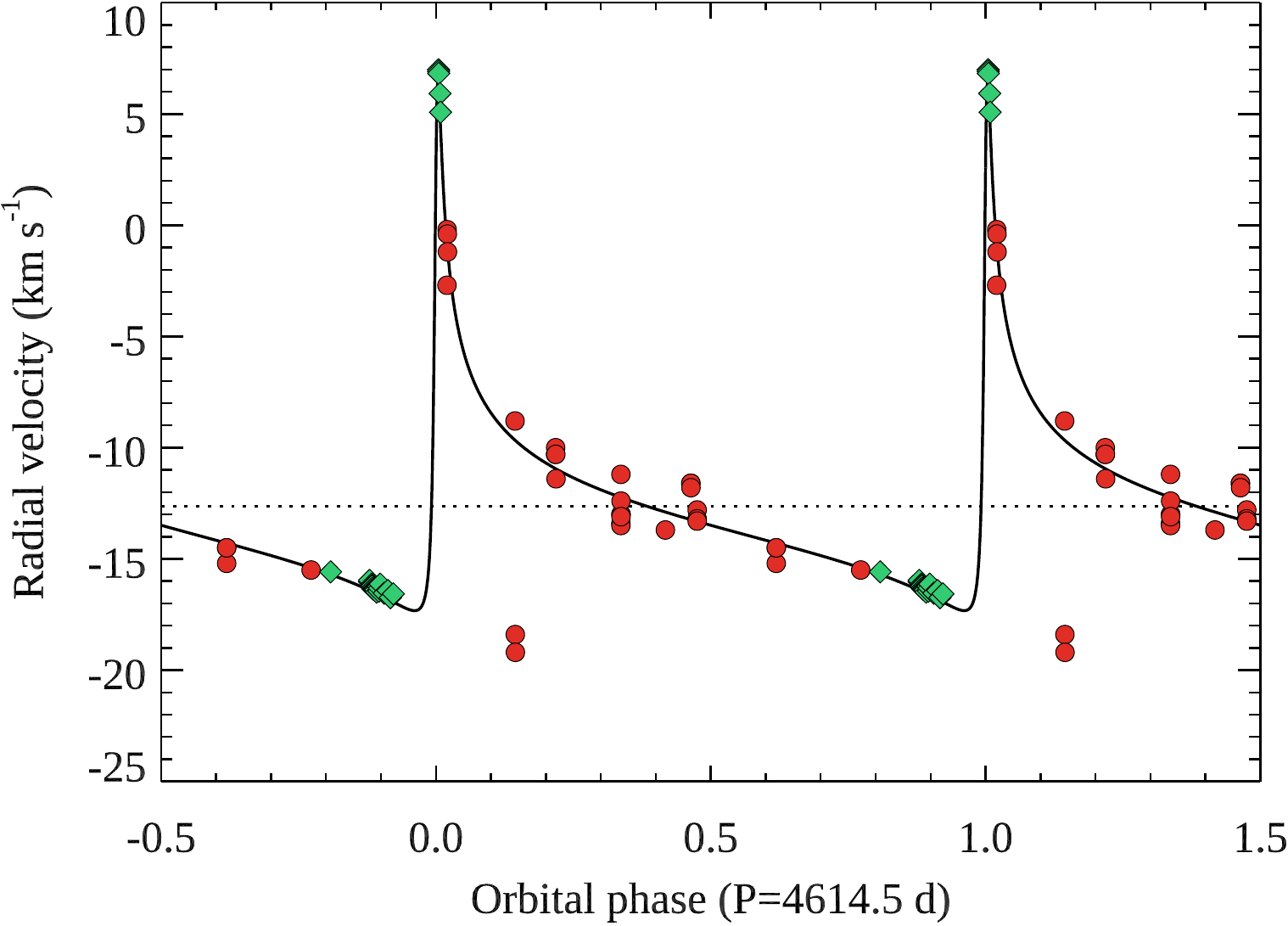}
\caption{Orbital radial velocity variation of Alhena. The solid line shows the orbital solution by \citet{lehmann02}. The red circles correspond to the measurements by \citet{adelman15} shifted by $-1.4$~km\,s$^{-1}$. The green diamonds are our measurements reported in Table~\ref{journal_alhena}.}
\label{rv_orbit}
\end{figure}

Alhena is a single-line spectroscopic binary in a long-period, highly eccentric orbit. \citet{scholz97} analyzed several hundred radial velocity measurements covering a time span of over 100 yr, establishing an orbital period of 4614.51~d (12.6~yr). Their orbital solution has been updated by \citet{lehmann02}, who derived $K_1$ = 11.9 km\,s$^{-1}$ and $e$ = $0.893$. Sixteen new radial velocity measurements were reported by \citet{adelman15} and 25 data points are provided by our study (Table~\ref{journal_alhena}). Radial velocity measurements were obtained by fitting the LSD I profiles with a double Gaussian function (enabling a convincing fit of both the core and wings of the line) with the same centroid.  In Fig.~\ref{rv_orbit} we compare the orbital solution derived by \citet{lehmann02} with the recent radial velocity measurements. The data from \citet{adelman15} had to be shifted by $-1.4$~km\,s$^{-1}$ to bring it into agreement with the published spectroscopic orbit. In addition, their two data points near the orbital phase of 0.15 are likely to be erroneous. On the other hand, our 25 measurements show an excellent agreement with the previous spectroscopic orbital solution.

\begin{figure}
\centering
\includegraphics[width=8cm]{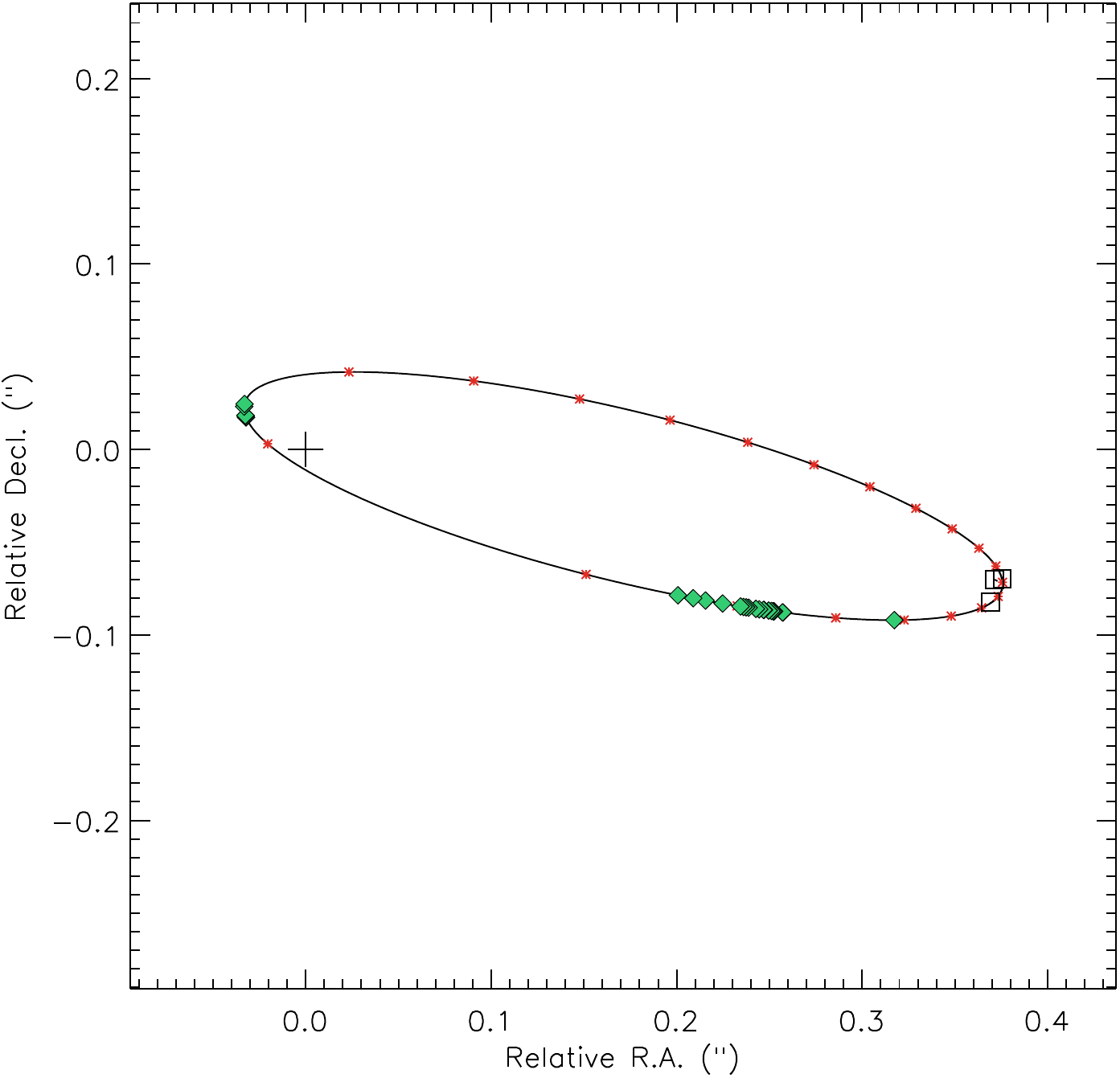}
\caption{Visual orbit of Alhena B around the primary (indicated by a cross) according to \citet{jancart05}, \citet{drummond14} and \citet{thalmann14}. The open square symbols show astrometric measurements from the latter two studies. The small red asterisks are plotted with a 0.05 step in phase. The green diamonds indicate phases of our spectropolarimetric observations of Alhena.}
\label{visual_orbit}
\end{figure}

In addition to the spectroscopic radial velocity variation discussed above, some information on the visual orbit is available. \citet{jancart05} derived the primary's orbit from the photocentre motion evident in the Hipparcos data. Later \citet{drummond14} was able to resolve the two components using adaptive optics observations at wavelengths around H$\alpha$. Two additional astrometric measurements were reported by \citet{thalmann14} based on observations in the H-band. According to his results, the orbital semi-major axis is 0.2732\arcsec, the orbital inclination is 107$^{\circ}$ and the secondary is about 3 magnitudes fainter than the primary. Using Hipparcos parallax, \citet{drummond14} calculated masses of $3.4\pm0.8$~$M_\odot$ and $1.4\pm0.3$~$M_\odot$ for the primary and secondary respectively, which is different from but compatible with the values provided by \cite{fekel93} (see Sect.~\ref{intro}). The visual orbit of Alhena is illustrated in Fig.~\ref{visual_orbit}. Our observations of this star were obtained at the orbital phases shortly before and just after the periastron passage, when the separation of the components decreased to about 1~AU.
%However, the very long orbital period of Alhena excludes any efficient tidal interactions between its components. Indeed, if we apply the theory for the dissipation of stellar tides developed by \cite{Zahn1975,Zahn1977} for early-type stars, where tidal gravity waves are damped through thermal diffusion, we obtain characteristic circularisation and synchronisation timescales for Alhena longer than the age of the Universe.

\section{Magnetic measurements}
\label{sect_mag}

We applied the Least-Squares Deconvolution (LSD) technique \citep{donati97} on each individual spectrum, in order to construct a single profile with an increased signal-to-noise, that allows us to detect weak Zeeman signatures in Stokes V profiles. To compute the LSD profiles, we use the same mask as B16, which contained 1052 spectral lines. This mask was extracted from VALD3 \citep{piskunov95, kupka99} using a $T_{\rm eff}$=9250 K and $\log$ g=3.5. The lines from the mask absent in the observed spectrum and the lines blended with hydrogen were removed from the mask. For each spectral line, the mask contains the wavelength, line depth, and Land\'e factor, to be used by the LSD method. The depths were adjusted to fit the depth of the observed spectral lines.

Thanks to this mask, we extracted LSD Stokes I and V profiles for each observation, as well as the null (N) polarization profiles to check for spurious signatures. For the nights when several spectra were taken, we co-added them to obtain a nightly-averaged LSD profile. An example of the LSD Stokes I, Stokes V and N profiles is shown in Fig.~\ref{lsd}. A plot of all profiles is available in the appendix (Fig.~\ref{lsd_annexe}). Zeeman signatures are clearly seen for all nights. All of them had a normal shape with a positive and negative lobe. For all observations, the N profiles contained only noise, indicating that the V profiles are not contaminated by spurious signal.

To evaluate the presence of significant signal in the Stokes V LSD profiles, we computed the false alarm probability (FAP) by comparing the signal inside the lines with no signal \citep{donati97}. To obtain a definite detection, we need a FAP $<$ 0.001\%; between 0.001\% $<$ FAP $<$ 0.1\%, we obtained a marginal detection, otherwise there is no detection. Table~\ref{bl_alhena} indicates what kind of detection, we obtain for each night: definite detection (DD), marginal detection (MD) or no detection. We obtained 23 DD and 2 MD.  

\begin{figure}
\includegraphics[scale=0.34, clip]{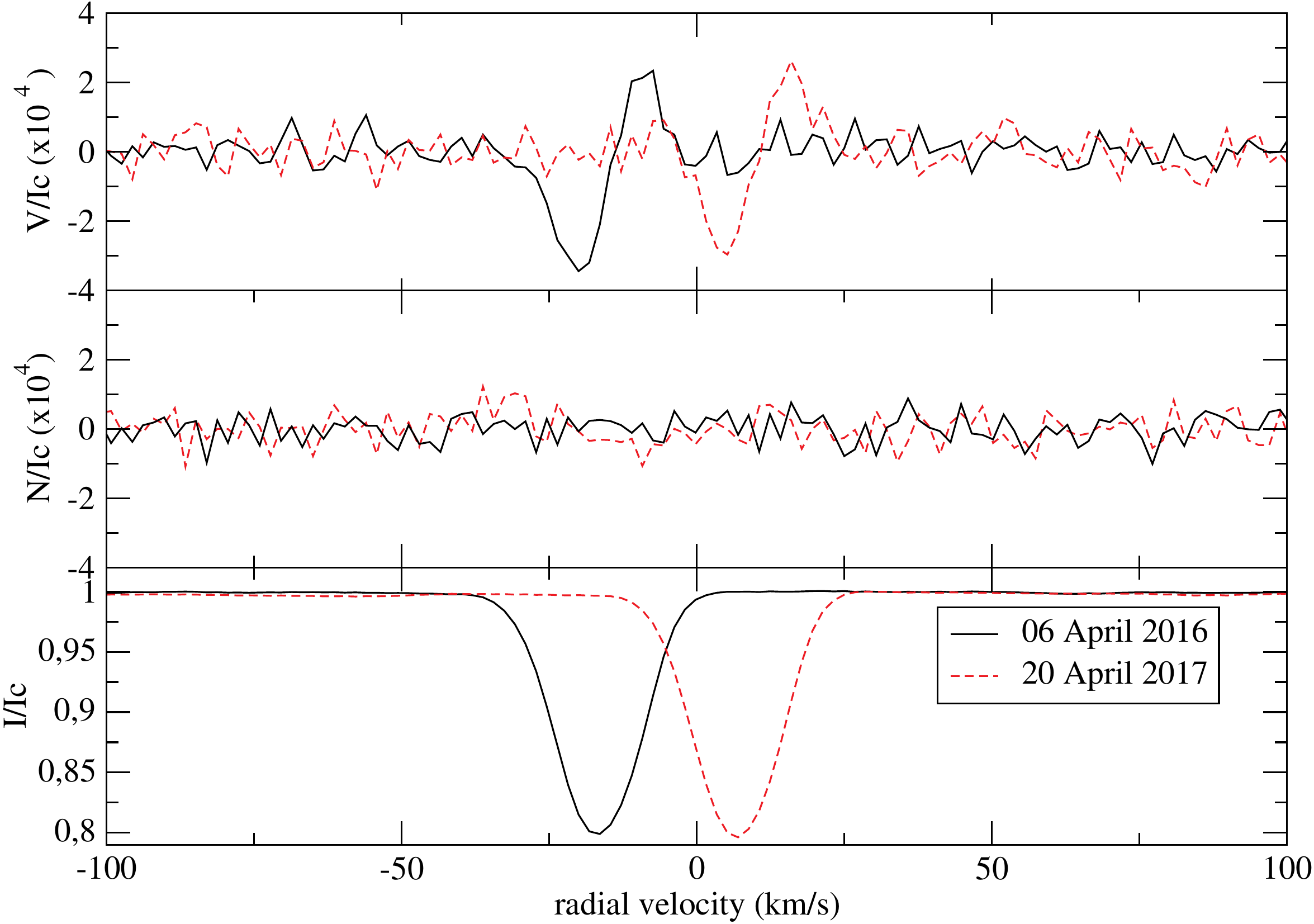}
\caption{Example of LSD Stokes I (bottom), Stokes V (top), and null N (middle) profiles for two different nights of observations.}
\label{lsd}
\end{figure}

 We calculated the longitudinal field value ($B_l$) corresponding to the magnetic signatures using the center-of-gravity method \citep{rees79} with a mean wavelength of 500 nm and a mean Land\'e factor of $\sim$1.46 corresponding to the normalization parameters used in the LSD code, and taken a velocity range of $\pm$21 \kms around the center of the line: 

\begin{equation}
B_{l} = -2.14\times10^{-11}\frac{\int v V(v)dv}{\lambda_{0}g_{m}c \int (1-I(v))dv} G.
\label{bl_measure}
\end{equation}

where $\lambda_{0}$ is the mean wavelength and  $g_{m}$ is the mean Land\'e factor.

For each night of observations, we reported the longitudinal magnetic field value, and the corresponding N values in Table~\ref{bl_alhena}.The N values listed in the same table were computed from N profiles using again equation \ref{bl_measure} with the same integration limits. The longitudinal magnetic field measurements are negative for all observations and vary between -10 G and -3 G, with typical error bars below 3 G. The values extracted from the N profiles are compatible with 0 G within 2$\sigma$N, where $\sigma$N is the error on N, for each night of observations (see Table ~\ref{bl_alhena}).

\begin{table}
\caption{Longitudinal magnetic field (B$_l$) of Alhena A and null (N) measurements with their respective error bars, magnetic detection status and the rotational phase}
\centering
\begin{tabular}{c c c c c}
\hline
\#  & B$_l \pm \sigma$B$_l$ (G) & N$\pm \sigma$N (G) & Detection & Rot. Phase\\
\hline
\hline
1 & -3.72$\pm$2.25 & -1.36 $\pm$2.25 & DD & 0.212\\
02 &  -6.50$\pm$2.33 	&  2.17$\pm$2.31 & DD & 0.499\\
03 & -6.58$\pm$2.12  &  -1.90$\pm$2.12 & DD & 0.610\\
04 & -10.34$\pm$2.06 &	 0.54$\pm$2.06 &  DD & 0.728\\
05 & -6.32$\pm$1.99  &  -2.03$\pm$1.99 & DD & 0.839\\
06 & -6.50$\pm$2.42	&  1.72$\pm$2.43 & DD & 0.949\\
07 &  -8.61$\pm$2.46  &  -2.83$\pm$2.43 & DD & 0.381\\
08 & -9.45$\pm$2.87  &  -1.94$\pm$2.85 & DD & 0.058\\
09 & -6.68$\pm$2.24	&  0.17$\pm$2.24 & DD & 0.177\\
10 & -5.43$\pm$2.06	&  3.32$\pm$2.04 & MD & 0.289\\
11 & -10.01$\pm$2.54  &  -1.77$\pm$2.52 & MD & 0.290\\
12 & -3.79$\pm$2.49	&  0.14$\pm$2.49 & DD & 0.059\\
13 & -4.53$\pm$2.47  &  -0.40$\pm$2.47 & DD & 0.726\\
14 & -6.65$\pm$1.79 &  0.15$\pm$1.79 & DD & 0.288\\
15 & -6.66$\pm$2.00  &  -2.08$\pm$2.02 & DD & 0.841\\
16 & -7.19$\pm$2.38	&  1.15$\pm$2.39 & DD & 0.506\\
17 & -8.08$\pm$1.05 & 1.03$\pm$ 1.05 & DD & 0.290\\
18 & -8.31$\pm$1.78 & 0.51$\pm$1.78 & DD & 0.722\\
19 & -5.15$\pm$0.98 & -0.16$\pm$0.98 & DD & 0.054\\
20 & -8.82$\pm$1.89 & 1.09$\pm$1.90 & DD & 0.829\\
21 & -5.15$\pm$1.40 & -2.14$\pm$1.40 & DD & 0.010\\
22 & -5.15$\pm$1.02 & 0.28$\pm$1.02 & DD & 0.122\\
23 & -6.12$\pm$1.11 & -0.25$\pm$1.11 & DD & 0.232\\
24 & -8.49$\pm$1.27 & -0.64$\pm$1.27 & DD & 0.458\\ 
25 & -5.59$\pm$1.05& 0.11$\pm$1.05 & DD & 0.902\\
\hline
\end{tabular}
\label{bl_alhena}
\end{table}

B16 considered that the detected magnetic field comes from the Am component of the binary system due to the fact that the brightness of the secondary is weak and its spectral contribution is negligible. The observations in 2017 are taken just after the periastron passage and, due to the high eccentricity of the binary ($e$=0.89), the LSD Stokes I profiles of the primary component are shifted (see Fig.~\ref{lsd}) compared to the ones obtained in 2015/2016. The signature in the Stokes V profile followed this shift in radial velocity, confirming that it is the primary (the Am star) that is magnetic.  

\section{Magnetic modeling}
\label{sect_field}

\subsection{Search for the rotation period}
\label{rot}

The first step to model the magnetic field of Alhena is to search for the star's rotation period. The magnetic fields of intermediate-mass stars are generally dominated by a strong dipolar component with a magnetic axis inclined with respect to the rotation axis. This configuration causes a rotational modulation of the longitudinal magnetic field. We searched for a rotation period in the 25 longitudinal magnetic field measurements of Alhena with the clean-NG algorithm \citep[see][]{gutierrezsoto09}. We did not find a significant period with this method, due to the small variations of the longitudinal magnetic field values compared to the measured uncertainties.

\begin{figure}
\includegraphics[scale=0.5, trim=2cm 0.5cm 0.5cm 1cm]{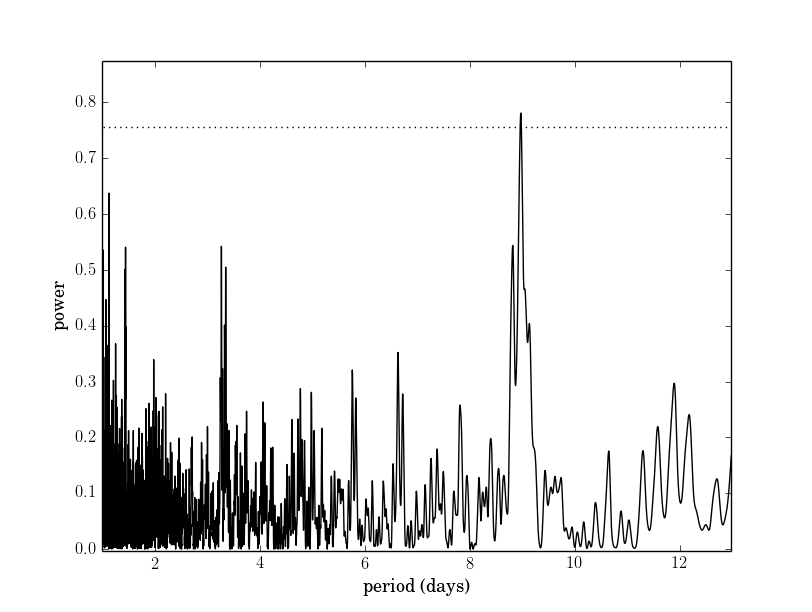}
\caption{The power spectrum of the equivalent width of the LSD Stokes I profiles of Alhena. The highest peak corresponds to the period $f$=8.975 days identified as the rotation period. The dashed line corresponds the 3$\sigma$ confidence level.}
\label{period}
\end{figure}

However, variability was observed in the LSD Stokes I profiles (see Fig.~\ref{LSDI}). This variability can be linked to stellar rotation. We therefore measured the equivalent width of each LSD profile and searched for a period thanks to the clean-NG algorithm \citep[see][]{gutierrezsoto09}. We obtained only one  significant period at 8.975$\pm$0.05 days (Fig.~\ref{period}). Figure \ref{EW} shows the variation of the equivalent width of the LSD I profile in phase with the period. We show in Fig.~\ref{EW} the difference between the mean LSD profile and each individual LSD I profiles in phase with this period. The variations are in good agreement with this period. These variations in the LSD Stokes I profiles correspond to either abundances spots or temperature spots at the surface of Alhena\,A. As a first attempt to investigate the origin of these spots, we obtained LSD profiles with different line masks restricted to the lines of single chemical species (e.g. Fe, Ti, Cr), in order to test if the observed variability is more prominent in specific species, as expected in the case of chemical spots. We then computed a similar dynamic plot than for the LSD profiles performed with all chemical species (these plots are not shown here). The outcome is a similar pattern for each individual chemical element than for the global LSD profiles. This result may be a first hint that the variations in the intensity profiles could be linked to temperature spots. Unfortunately, the relatively small projected rotational velocity of Alhena makes it a difficult target for Doppler Imaging  (because the signature of spots in Stokes I become smaller than inaccuracies in our simplified model of the average line profile. This will not be the case in Stokes V, were the average line has a null depth), preventing us from mapping the surface distribution of these inhomogeneities.

\begin{figure}
\includegraphics[scale=0.36,clip]{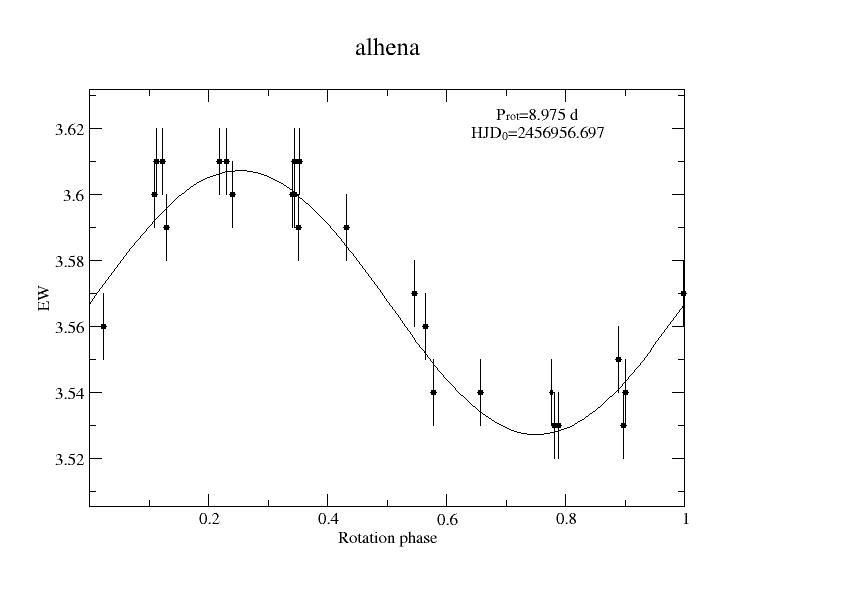}
\caption{Rotational modulation of the equivalent width of the LSD profiles of Alhena\,A. }
\label{EW}
\end{figure}

\begin{figure}
\includegraphics[scale=0.27,clip]{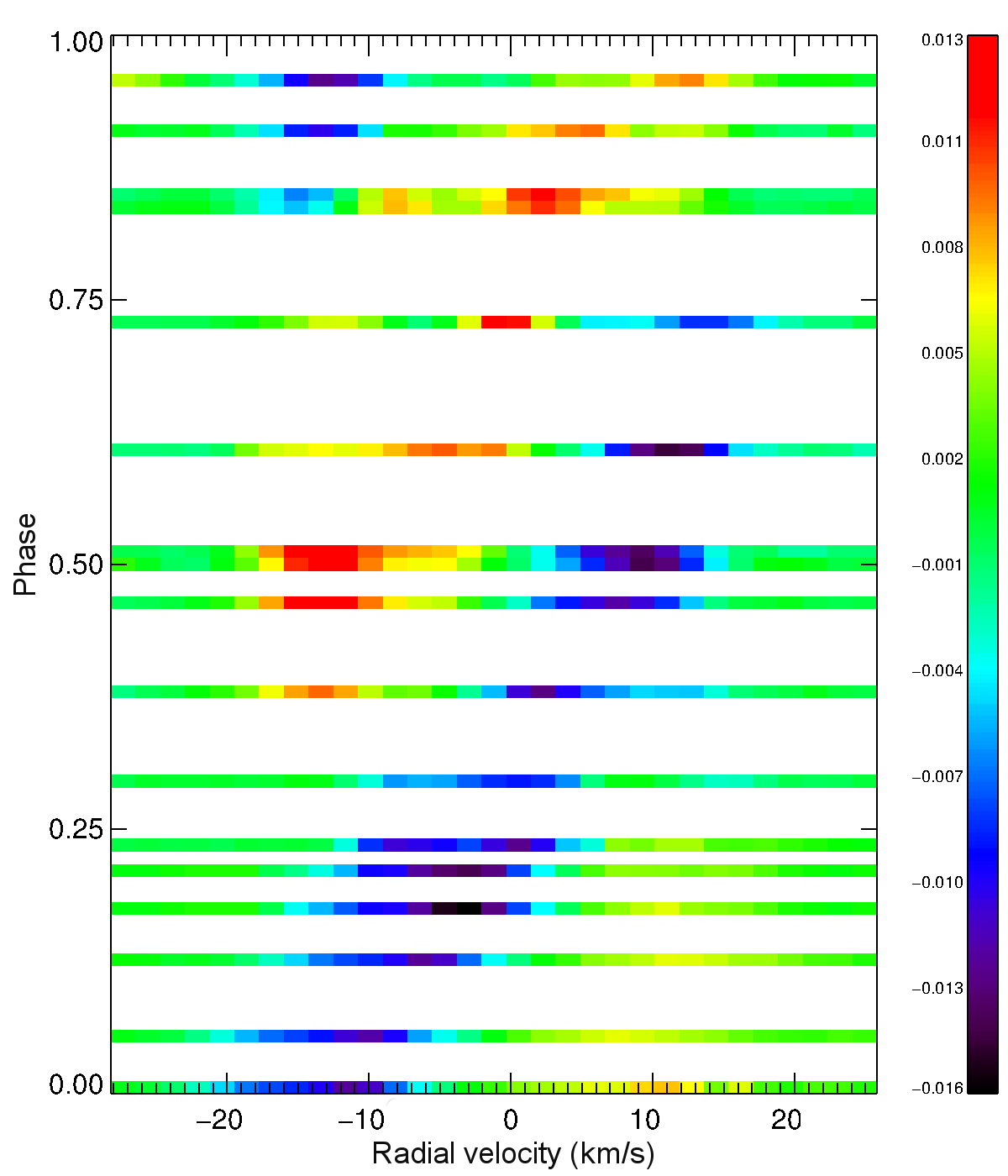}
\caption{Dynamic plot of the difference between the mean LSD Stokes I profiles and the individual LSD Stokes I profiles of Alhena\,A, folded with the rotation period of 8.975 days.}
\label{EW}
\end{figure}

\begin{figure}
\includegraphics[scale=0.36, trim= 0cm 0cm 0cm 1cm,clip]{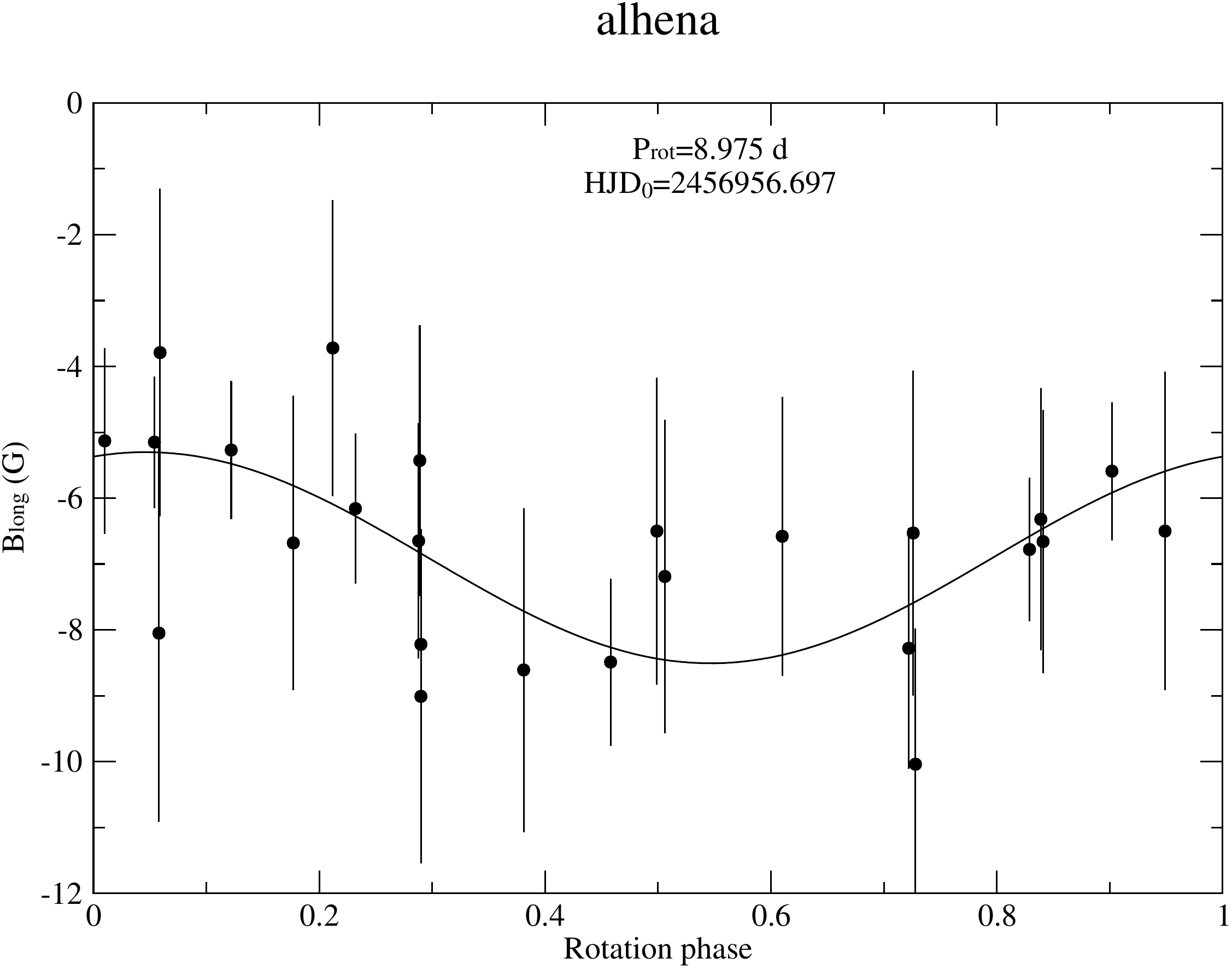}
\caption{Rotational modulation of the longitudinal magnetic field of Alhena\,A. The black line corresponds to the best dipolar fit. }
\label{modulation_rot}
\end{figure}

\subsection{Longitudinal magnetic field modeling}
\label{longitudinalField}

 We calculated the inclination angle of Alhena A, using the period detected in Sect.~\ref{rot} as rotational period of the star, a radius of 4.03$\pm$0.1 R$_\odot$ \citep{royer07}, and a projected rotational velocity \vsini\  of 10.74$\pm$ 0.2 \kms \citep{gray14}. We obtain an inclination angle $i$ = 28.2 $\pm$ 1.6$^\circ$. 
Assuming that the longitudinal field experiences sinusoidal variations following the period of Sect.~\ref{rot} (Fig.~\ref{modulation_rot}), we can deduce the obliquity angle $\beta$ of the magnetic field with respect to the rotation axis. For that, we used the formula $r = B_{\rm min} / B_{\rm max} = \cos$($i - \beta$)$ / \cos$($i $+$ \beta$) \citep{preston67}. Using the outcome of the sinusoidal fit, we obtain $r$=0.73$\pm$0.22 and $\beta$ = 16.6 $\pm$ 14 $^\circ$. 

In addition, we can estimate the polar field strength, using the obliquity $\beta$ and the inclination $i$ determined above, with the formula:

\begin{equation}
B_{\rm d} =B_{\rm max} \left( \frac{15+u}{20(3-u)}(cos\beta cos i + sin\beta sin i) \right) ^{-1}
\end{equation}

where the limb-darkening coefficient $u$ is assumed to be 0.3 \citep{claret04}. We found $B_{\rm d}$ = 26.6$\pm$4.9 G.

\subsection{Stokes V modeling}
\label{directModeling}

The Stokes V signatures was modeled using the oblique rotator model, assuming  a dipole and a rotational period of 8.975 days.

The five free parameters of the model are the inclination $i$, the obliquity angle $\beta$, the dipolar magnetic field strength B$_d$, a phase shift $\phi$ and the off-centering distance d$_d$ of the dipole with respect to the center of the star (d$_d$=0 for a centered dipole and d$_d$=1 if the center of the dipole is at the surface of the star).

\begin{figure*}
\includegraphics[scale=0.75, trim= 1cm 0.5cm 0.5cm 1cm]{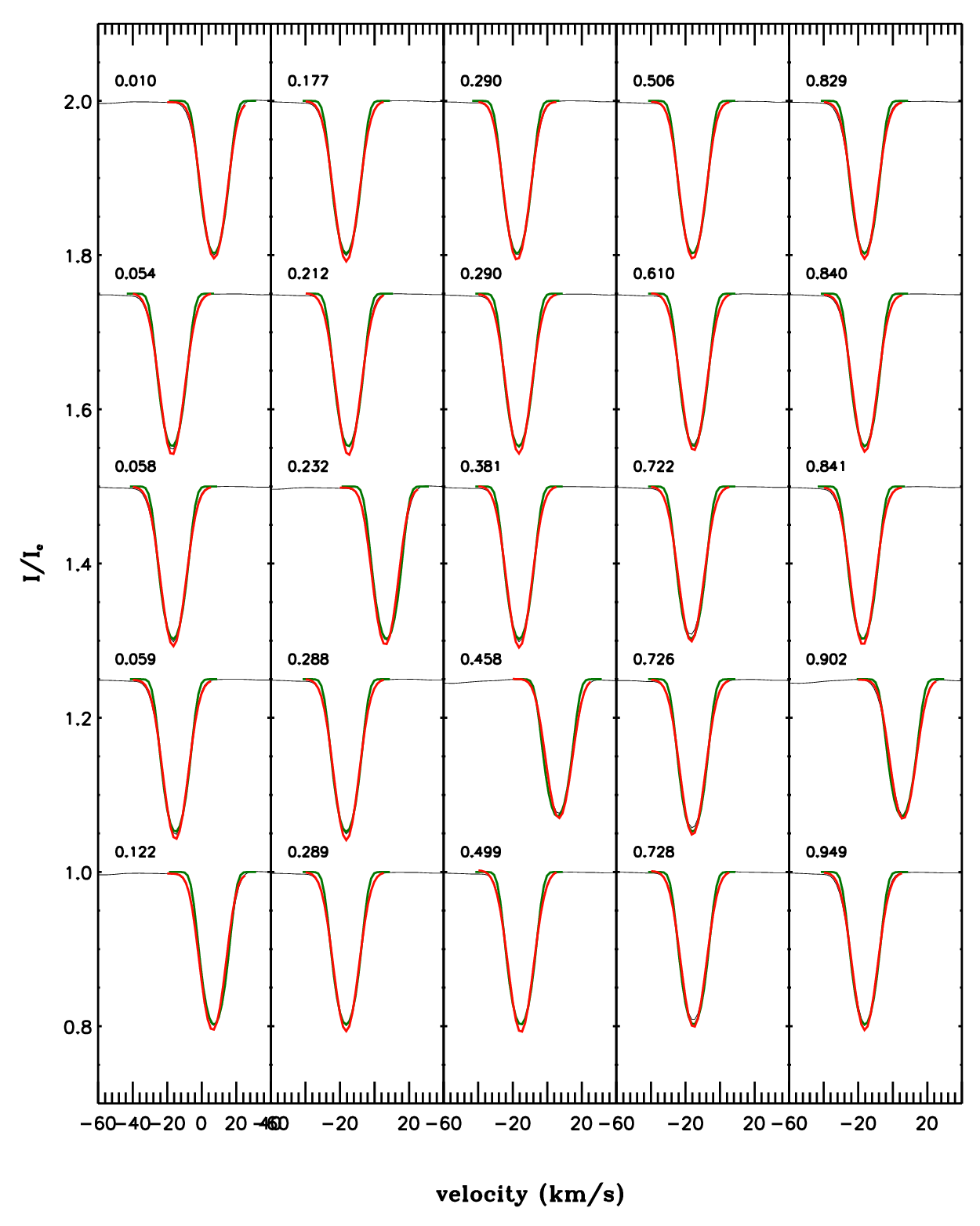}
\caption{Dipolar model fit (red) and ZDI fit (green) of the observed Stokes I profiles (black) of Alhena A, with vertical shifts for display purpose. The numbers next to every profile correspond to their rotational phase.} 
\label{I_model}
\end{figure*}

We used Gaussian local intensity profiles, with a width calculated according to the resolving power of NARVAL and a macro-turbulence of 6.1 \kms \citep{gray14}. The depth of local intensity profiles was calculated by fitting the observed LSD I profiles (see Fig. \ref{I_model}), and the adopted \vsini\ value (consistent with  \citealt{gray14}) is the one providing us with the best fit to the line width. Radial velocities are taken from Tab. \ref{journal_alhena}. We used the weighted mean Land\'e factor and wavelength derived from the LSD mask applied to the Narval observations and the rotation period of 8.975 days to compute the synthetic Stokes V profiles. 

\begin{figure*}
\includegraphics[scale=0.75, trim= 1cm 0.5cm 0.5cm 1cm]{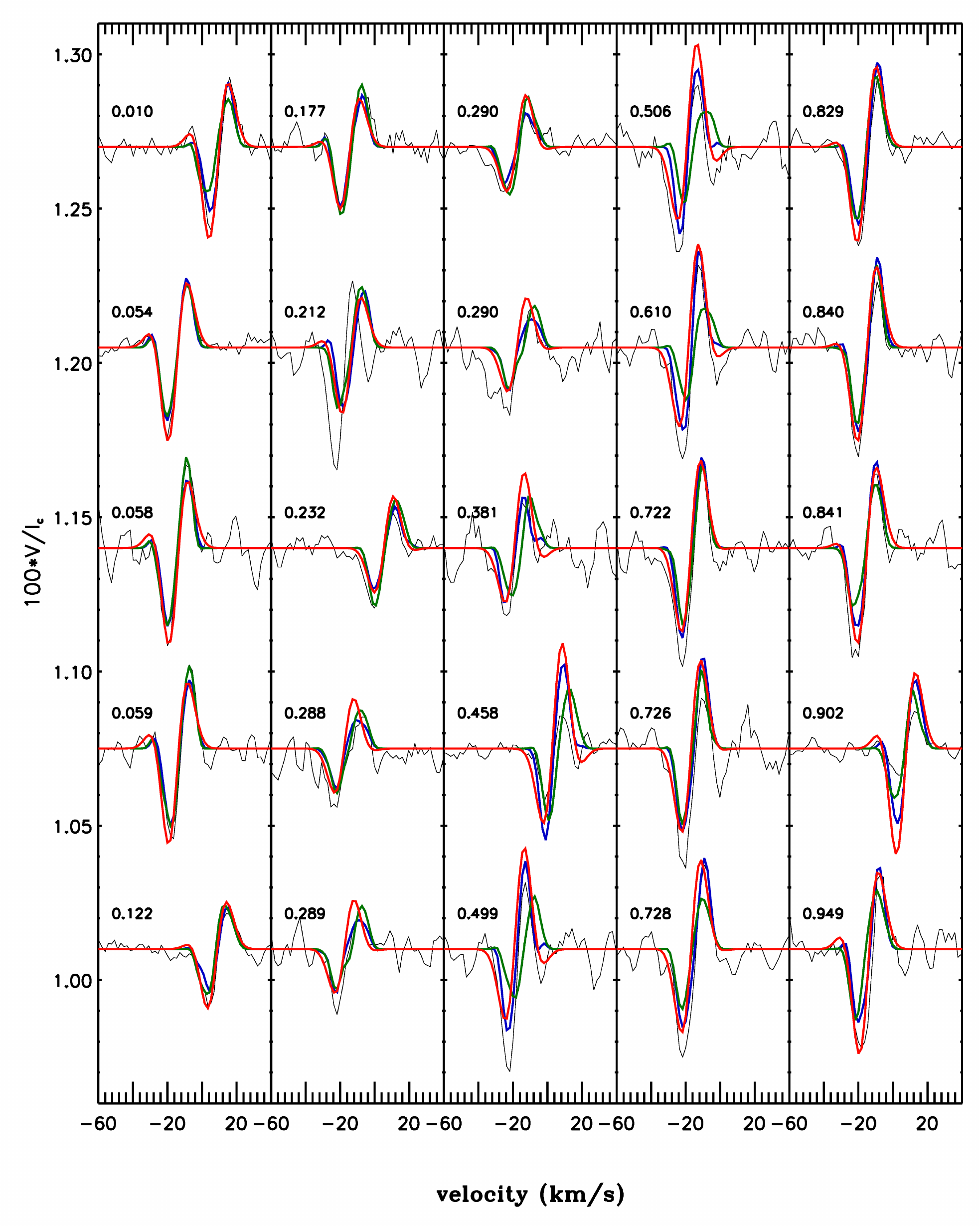}
\caption{Best dipolar model fit (red), best ZDI fit (green) and best ZDI fit including differential rotation (blue) of the observed Stokes V profiles (black) of Alhena A (with vertical shifts for display purpose). The numbers next to every profile correspond to their rotational phase.} 
\label{V_model}
\end{figure*}

We computed a grid of synthetic Stokes V profiles for each phase of observation by varying the five parameters mentioned above and applied a  minimization to obtain the best fit of all observations simultaneously. See \cite{alecian08} for  more details of the modeling technique. The parameters of the best fit ($\chi^2$=1.409) are $i$=22.8$\pm$4.1$^\circ$, $\beta$=34.1$\pm$3.4$^\circ$, B$_{\rm d}$=32.4$\pm$1.7G and d$_d$=0.007$\pm$0.014, where the error bars correspond to a 3$\sigma$ confidence level.  The off-centering distance is compatible with 0 and implies that the magnetic field of Alhena A is consistent with a dipole at the center of the star. 

Moreover, Fig.~\ref{V_model} shows the comparison between the observed and the best synthetic LSD V profiles for all observations. The model fits quite well the Stokes V profiles, however it is not perfect. The difference between the model and the observations suggests that the structure of the magnetic field is more complex than a dipole or that the magnetic field geometry has evolved over the course of the follow-up observing campaign, spread over $\sim 2.5$~yr. 

To resolve these discrepancies, we proceed with Zeeman-Doppler Imaging of Alhena. 

\subsection{Zeeman-Doppler Imaging}

\begin{figure*}
\mbox{
\includegraphics[width=6cm]{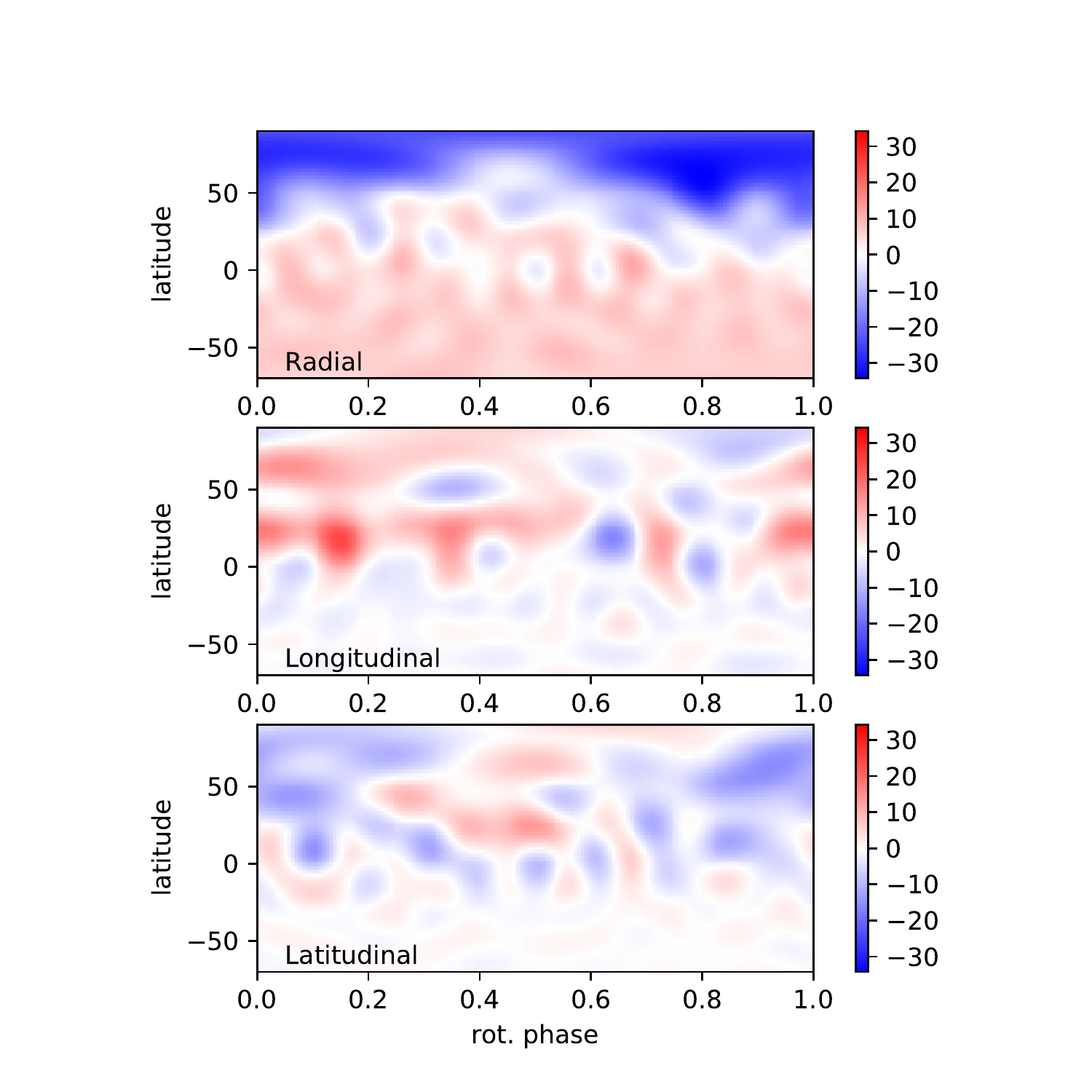}
\includegraphics[width=6cm]{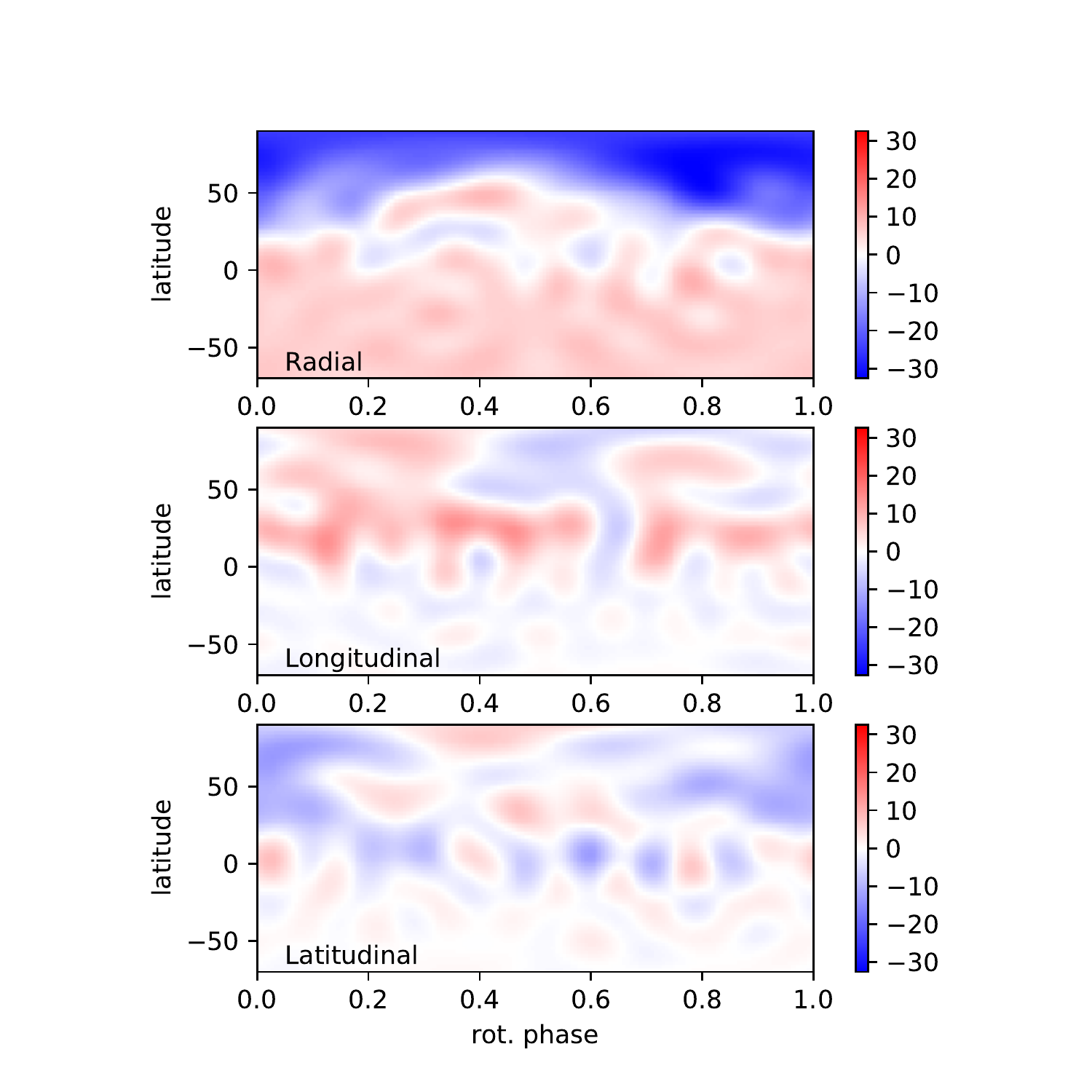}
\includegraphics[width=6cm]{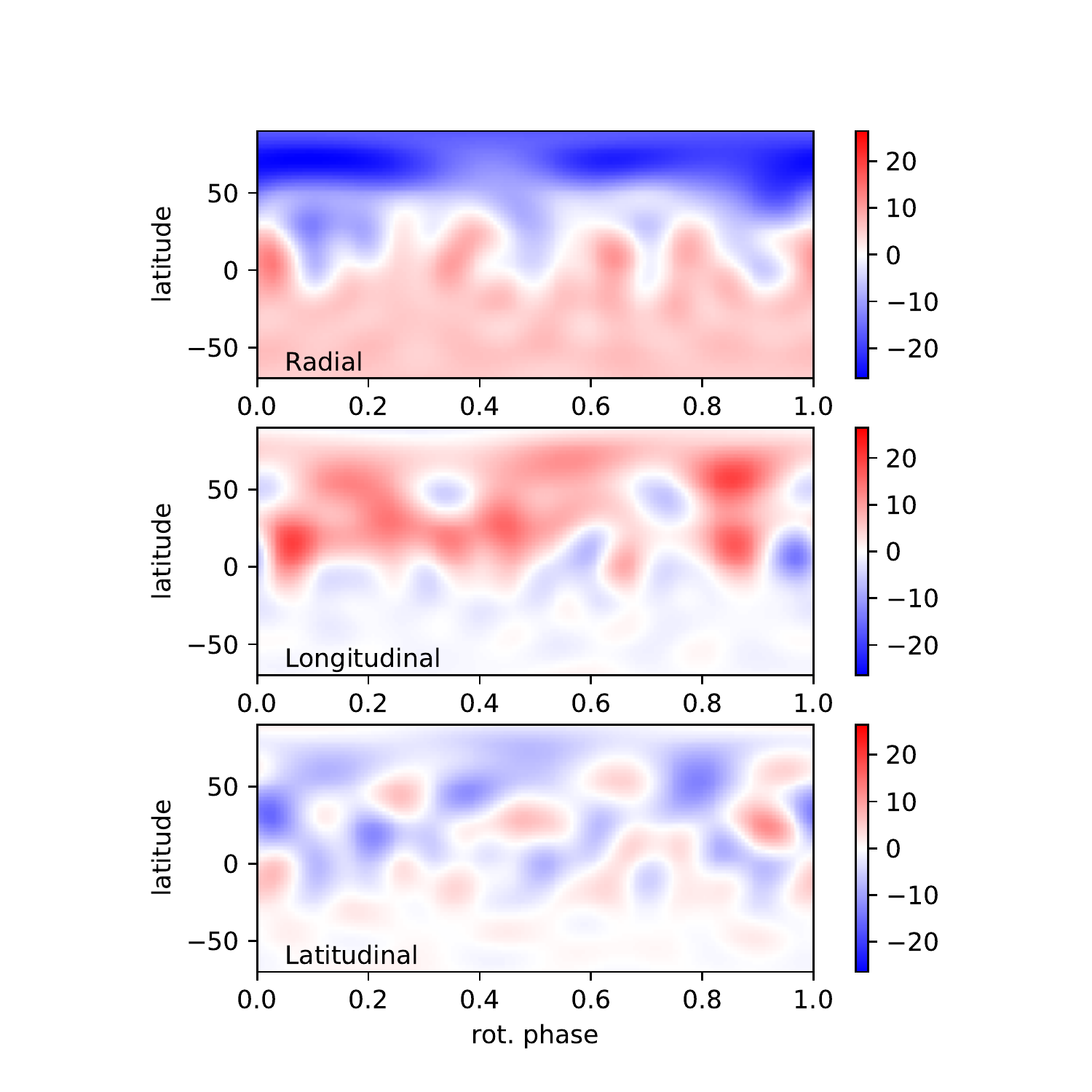}
}
\caption{ Magnetic geometry of Alhena obtained from different ZDI models. The radial, azimuthal, and meridional components of the magnetic field vector are plotted from top to bottom, in equatorial projection. The color scale illustrates the field strength, in Gauss. The left panel shows the map obtained from the 2015-2016 subset, with differential rotation. The middle panel is derived from the full data set (2014-2017), with differential rotation. The right panel is also for the full data set, but assuming solid-body rotation (note the different color scale for this panel).}
\label{magneticGeometry}
\end{figure*}

Tomographic mapping offers another method to model the surface magnetic topology of the star, allowing for the reconstruction of complex field structures. We employ the Zeeman-Doppler Imaging method (thereafter ZDI, \citealt{semel89}), with a spherical harmonics decomposition of the field topology \citep{donati06}. We use here a recent Python implementation of this algorithm, described by \cite{folsom18}.

The magnetic model underlying ZDI assumes that the stellar surface is paved with rectangular pixels having roughly the same area. A magnetic vector is attributed to every pixel, with a given strength and orientation, and under the constraint that the overall field geometry is described by a sum of spherical harmonics modes. A local Stokes I and V pseudo-line profile is also attributed to the pixel. Locally, Stokes I is simply modeled by a Gaussian function of adjustable depth and width, while Stokes V is computed under the weak field approximation (and is therefore proportional to the derivative of Stokes I). For a given rotational phase, local Stokes I and V profiles located on the visible hemisphere of the stellar surface are weighted according to a projection factor and a linear limb-darkening law, and are shifted in wavelength according to the local Doppler velocity.  The wavelength and Land\'e factor of the pseudo-line is chosen identical to the values given in Sec. \ref{directModeling}, as well as the \vsini. The depth of the local Stokes I profiles is tuned so that the depth of the integrated line profile matches the depth of the observed LSD profiles. The width of the Gaussian line (which includes any non-rotational broadening) is set to $2 \times 10^{-3}$~nm. This value of the width provides us with a convincing fit of the core of the line, although the extremity of the line wings are not correctly matched (Fig. \ref{I_model}). This limitation of our line model is however considered acceptable here, since observed Stokes V line profiles in the line wings stay below the noise level. The radial velocities used for ZDI are estimated independently (using a gaussian fit to the Stokes I line profiles), and provide us with estimates consistent with those listed in Tab. \ref{journal_alhena}. The equivalent width fluctuations reported in Sec. \ref{rot} were considered sufficiently small that their impact on the magnetic modelling is negligible. They are therefore not corrected here. This set of input parameters is used for all ZDI models described below.

From the whole available dataset, we  first restrict the tomographic reconstruction to a sub-time series running from 18 Sep. 2015 to 06 Apr. 2016. This subset consisted of  27 observations, which in practice sample 19 different rotation phases (given the repeated daily observations gathered starting on 20 Jan. 2016). Assuming a rotation period equal to 8.975~d (Sec. \ref{rot}), this dataset is free from any significant phase gap. 

The best ZDI model using this value of the rotation period (and assuming solid body rotation) is obtained for \vsini\ equal to $12\pm1$~km\,s$^{-1}$, slightly larger but within $2\sigma$ of the value adopted for direct Stokes V modeling (see Sect.~\ref{directModeling}). The optimal value for the inclination angle is $i$=$35\pm10^{\circ}$, again within $2\sigma$ of the values deduced in Sect.~\ref{directModeling} and \ref{longitudinalField}. Spherical harmonics modes up to $\ell_{\rm max} $=$ 10$ are allowed in the reconstruction. In practice the $\chi^2$ can be further reduced by increasing $\ell_{\rm max}$ to about 20, but the smaller magnetic structures showing up in the higher resolution maps display the typical pattern obtained when over-fitting the data. This $\ell_{\rm max}$ value leads to a total of 390 spherical harmonics coefficients to be adjusted. The number of pixels on the magnetic map has to be significantly larger than the number of spherical harmonics coefficients, and we adopted a minimal number of 3,000 pixels in our models. Our set of parameters leads to a reduced $\chi^2$ equal to 1.13, and the modeled Stokes V profiles are shown in Fig. \ref{V_model}.  The choice of other subsets (or even the complete data set) leads to similar results.

The main structure of the reconstructed magnetic geometry (Fig. \ref{magneticGeometry}) is a dipole, with the visible fraction of the stellar photosphere dominated by the negative pole featuring a polar field strength of about $36 \pm 4$~G  (Tab. \ref{tab:magE}, where uncertainties are derived similarly to \citealt{petit08}.). This value is compatible (to within $2\sigma$) with the one found with the oblique rotator model. The magnetic energy stored in the dipole alone amounts to $52 \pm 9$\% of the total magnetic energy of the poloidal field component, while the quadrupole and octupole account for $19\pm 2$\% and $6\pm 2$\% of the energy of the poloidal component, respectively. About 60\% of the total magnetic energy shows up in the axisymmetric field component, {\em i.e.} in spherical harmonics modes with $m$ = $0$. A toroidal field component is also reconstructed, containing $18\pm 5$\% of the total magnetic energy.

\begin{table*}
    \centering
    \caption{Magnetic properties of Alhena, derived from different ZDI models with solid body rotation ("solid" models) or differential rotation ("sheared" models). The data sets are indicated in the first line.
    We include the average and peak magnetic fields, $\langle B \rangle$ and $|B_{peak}|$ ($\langle B \rangle$ being also given for low order modes ($\ell < 4$)). We then list the fraction of magnetic energy in the toroidal component, the fraction of magnetic energy in the axisymmetric component (i.e. stored in spherical harmonics modes with $\ell = 0$), the fraction of magnetic energy in the dipole ($\ell=1$), quadrupole ($\ell=2$) and octopole ($\ell=3$) expressed as a fraction of the poloidal field energy.}
    \begin{tabular}{l|l|l|l|l}
        \hline
        Years & 2015-2016 & 2015-2016 & 2014-2017 & 2014-2017 \\
        ZDI model & solid   & sheared & sheared & solid \\
        \hline
        $\langle B \rangle$  & $10 \pm 1$ G & $10 \pm 1$ G & $9 \pm 1$ G & $9 \pm 1$ G  \\
        $\langle B \rangle$ ($\ell < 4$)  & $9 \pm 1$ G & $8 \pm 1$ G & $8 \pm 1$ G & $8 \pm 1$ G  \\
        $|B_{peak}|$  & $36 \pm 4$ G & $35 \pm 3$ G & $34 \pm 3$ G & $28 \pm 2$ G \\
        Toroidal (/total) & $18 \pm 5$  \%   & $17 \pm 4$ \% & $14 \pm 5$ \% & $22 \pm 6$ \% \\
        axisymmetric (/total)  & $61 \pm 6$ \% & $59 \pm 7$ \%  & $65 \pm 6$ \% & $64 \pm 7$ \%\\
        dipole (/poloidal)  & $52 \pm 9$ \% & $52 \pm 8$ \%  & $50 \pm 8$ \% & $46 \pm 10$ \%\\
        quadrupole (/poloidal)  & $19 \pm 2$ \% & $19 \pm 2$ \%  & $21 \pm 2$ \% & $15 \pm 2$ \%\\
        octopole (/poloidal)  & $6 \pm 2$ \% & $6 \pm 2$ \%  & $10 \pm 1$ \% & $7 \pm 2$ \%\\
        \hline
        
    \end{tabular}
    \label{tab:magE}
\end{table*}

\subsection{Surface differential rotation}
\label{sec:diffrot}

\begin{figure}
\includegraphics[width=9cm]{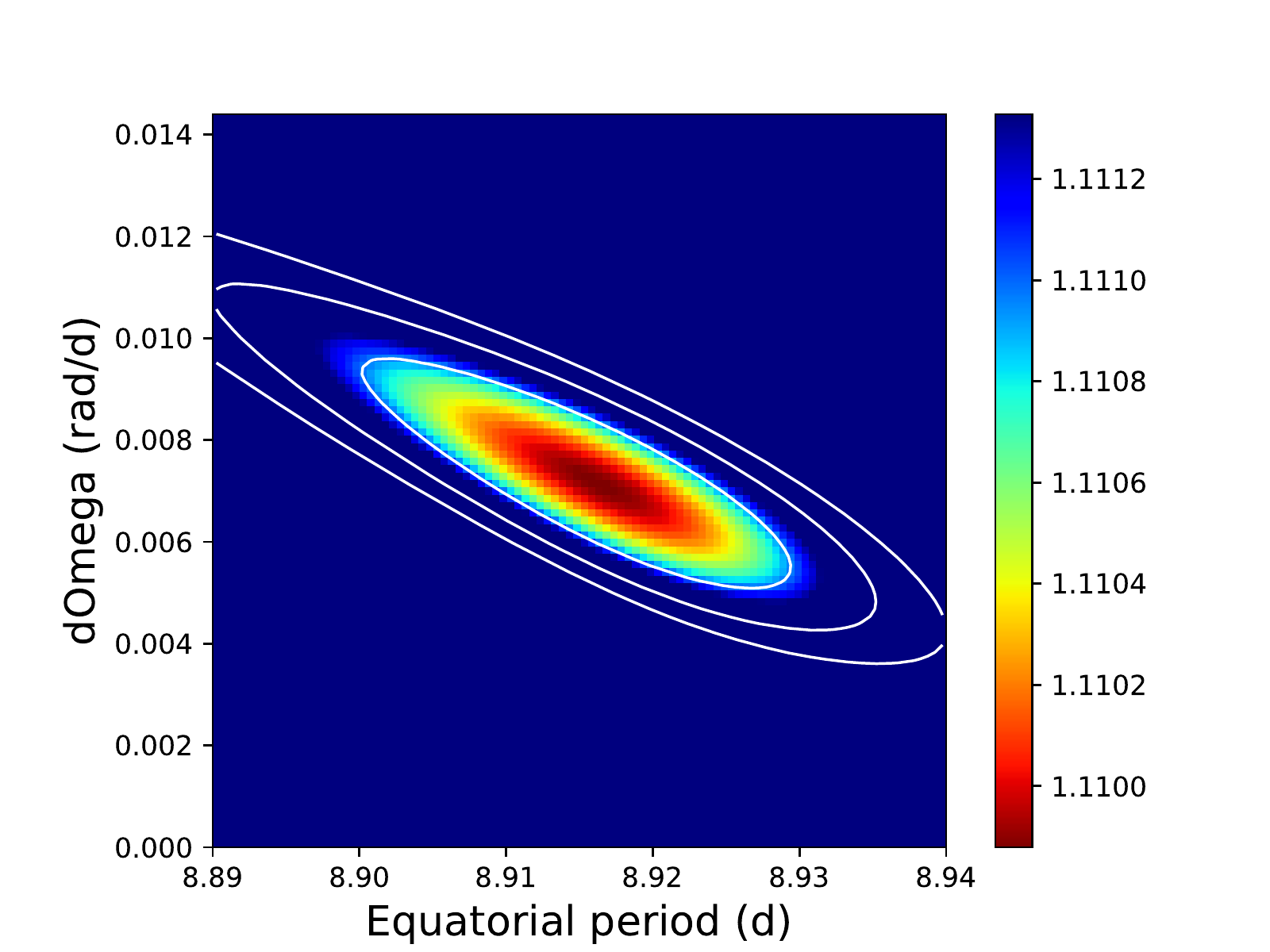}

\caption{$\chi^2$ landscape obtained for a grid of ZDI models using different values of the differential rotation parameters in Eq.~\ref{eq:diffrot}. A positive value of $d\Omega$, meaning a faster rotation of low latitudes, provides us with the best model. The color scale is set to saturate above a reduced $\chi^2$ equal to 1.1113, and the three white concentric lines show the $1\sigma$, $2\sigma$, and $3\sigma$ contours.}
\label{chi2Landscape}
\end{figure}

The fit accuracy provided by a direct Stokes V modeling (Sect.~\ref{directModeling}) was not able to reach the noise level, as shown by the reduced $\chi^2$ greater than one. The above ZDI model improved this situation by allowing for a more complex field topology, including a toroidal component. And yet, a small but noticeable mismatch remains. As an attempt to further improve our magnetic model, we now allow for progressive surface changes to modify the field geometry within the timespan of the data collection, in the form of a latitudinal differential rotation.

We use the fact that the rectangular pixels paving our synthetic star are distributed over latitudinal rings. It is possible to assume a different rotation period for every ring, so that the surface of the artificial star gets progressively distorted.  Each point on the surface is threaded by a magnetic field vector that is frozen for the entire timespan covered by the observations that are used to reconstruct the ZDI maps; as different regions of the surface rotate with different angular velocities, each vector is attached to its corresponding point on the surface such that the magnetic configuration changes over time.Here, we assume a smooth variation of the rotation rate as a function of the latitude, in the form:

\begin{equation}
\Omega (l) = \Omega_{\rm eq} - d\Omega.\sin^2(l)
\label{eq:diffrot}
\end{equation}

\noindent where $\Omega (l)$ is the rotation rate at stellar latitude $l$, $\Omega_{\rm eq}$ the rotation rate of the equator, and $d\Omega$ the difference in rotation rate between the equator and polar regions. Given that the changes in the shape of Stokes V line profiles owing to differential rotation are very subtle and difficult to detect, this solar-like differential rotation law is a simple way to model a systematic difference in rotation rate between magnetic features anchored at high versus low latitudes. 

Still using the 2015-2016 data subset, we computed a large number of magnetic geometries for different values of the free parameter doublet $(\Omega_{\rm eq}, d\Omega)$, and determined the doublet value that provides us with the best ZDI model assuming $\ell_{\rm max} = 10$ \citep{petit02}. $\chi^2$ variations around the $\chi^2$ minimum were used to derive statistical error bars on $\Omega_{\rm eq}$ and $d\Omega$ \citep{press92}. This method is now widely used for solar-type stars, where it proved to be a powerful tool to detect and quantify surface shears (e.g. \citealt{barnes05, petit08, morgenthaler12}). It was also applied to observations of hotter stars, but so far it was only able to confirm solid rotation states in these more massive objects \citep{donati06, briquet16}.

The outcome for Alhena is illustrated in  Fig.~\ref{chi2Landscape}, where a $\chi^2$ minimum is identified in the $\Omega_{\rm eq} - d\Omega$ plane. Note that a wider parameter space was probed as well using larger bins and including anti-solar shear values, without revealing any other local minimum at a similar $\chi^2$ level. The best model was obtained for $P_{\rm eq}$ = $8.915 \pm 0.015$~d and $d\Omega$ = $0.0073 \pm 0.0023$~rad.d$^{-1}$. The positive value for $d\Omega$ suggests that equatorial latitudes rotate faster than polar regions, while its level would indicate that the shear experienced by Alhena is about seven times weaker than observed on the Sun. According to our measurement, it takes about 860~d for the equator to lap the pole, versus about 120~d for the Sun. The rotation period near the pole is equal to 9.0~d, and the period of 8.975~d reported from Stokes I variability (Sec. \ref{rot}) is the one that is expected at a latitude of 53$\degr$.   

The map itself is only slightly affected by the inclusion of differential rotation in the model (left panel of Fig. \ref{magneticGeometry}). The new $\chi^2$ value is equal to 1.11. The remaining deviation from $\chi^2$=1 also shows that this refinement in the ZDI model is still unable to provide us with an optimal fit to our time series of observations. An inspection of Fig. \ref{V_model} suggests that the missing ingredient is a small level of asymmetry in Stokes V signatures. Observed Stokes V profiles seem to display a negative (bluewards) lobe systematically deeper than the positive (redwards) one. Our model does not feature such asymmetry, as it is not produced through the standard Zeeman effect. It is rather expected to be the product of a combination of vertical gradients in velocities and magnetic fields \citep{lopez02}. The asymmetry level is, in fact, quite limited for Alhena, at least compared to the extreme cases of Sirius~A, $\theta$~Leo and $\beta$~UMa \citep{petit11,blazere16a}.   

As stressed already, the timespan of our dataset (restricted here to observations gathered from 18 Sep. 2015 to 06 Apr. 2016) is significantly smaller than the lap time of differential rotation. This  justifies, {\em a posteriori}, our choice to base our analysis on this subset (since our crude description of a surface shear can lead to unrealistic estimates of the shear level whenever the surface of the artificial star is stretched over a time much longer than the lap time). We now consider the full data set (from 2014 to 2017) and reconstruct a new map assuming the same shear value (Fig. \ref{magneticGeometry} and Tab. \ref{tab:magE}). The outcome is very similar to the model obtained from the shorter subset. The $\chi^2$ value is equal to 1.1, the general aspect of the map is very close to the previous one (the most noticeable differences show up in small structures, especially in the low latitude azimuthal field component). The magnetic quantities of Tab. \ref{tab:magE} are also very close to the previous model. We then reconstruct another map from all available data, but this time assuming solid body rotation. The higher $\chi^2$ value (equal to 1.65) shows that this simpler model is not as successful at fitting the data, and this poor data adjustment is visible in Fig. \ref{V_model}. This time, the map looks different, with a more axisymmetric and less contrasted magnetic field. This is reflected in Tab. \ref{tab:magE}, where the average field stays close to values obtained with the other models, but with a smaller peak value (presumably showing that ignoring the surface shear has the effect to blur the map, since individual magnetic regions drift over the timespan of data collection).

\begin{figure}
\includegraphics[width=9cm]{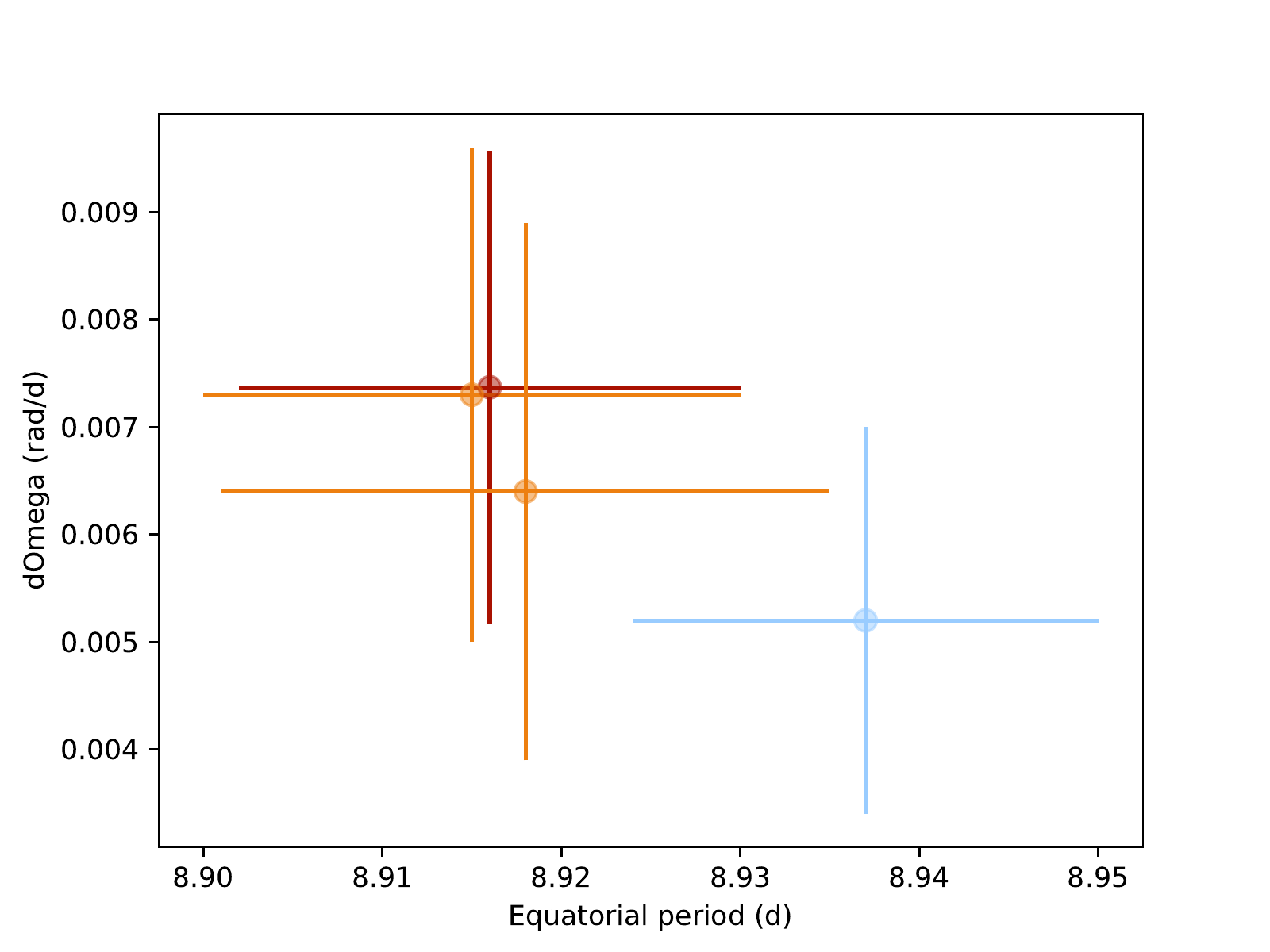}
\includegraphics[width=9cm]{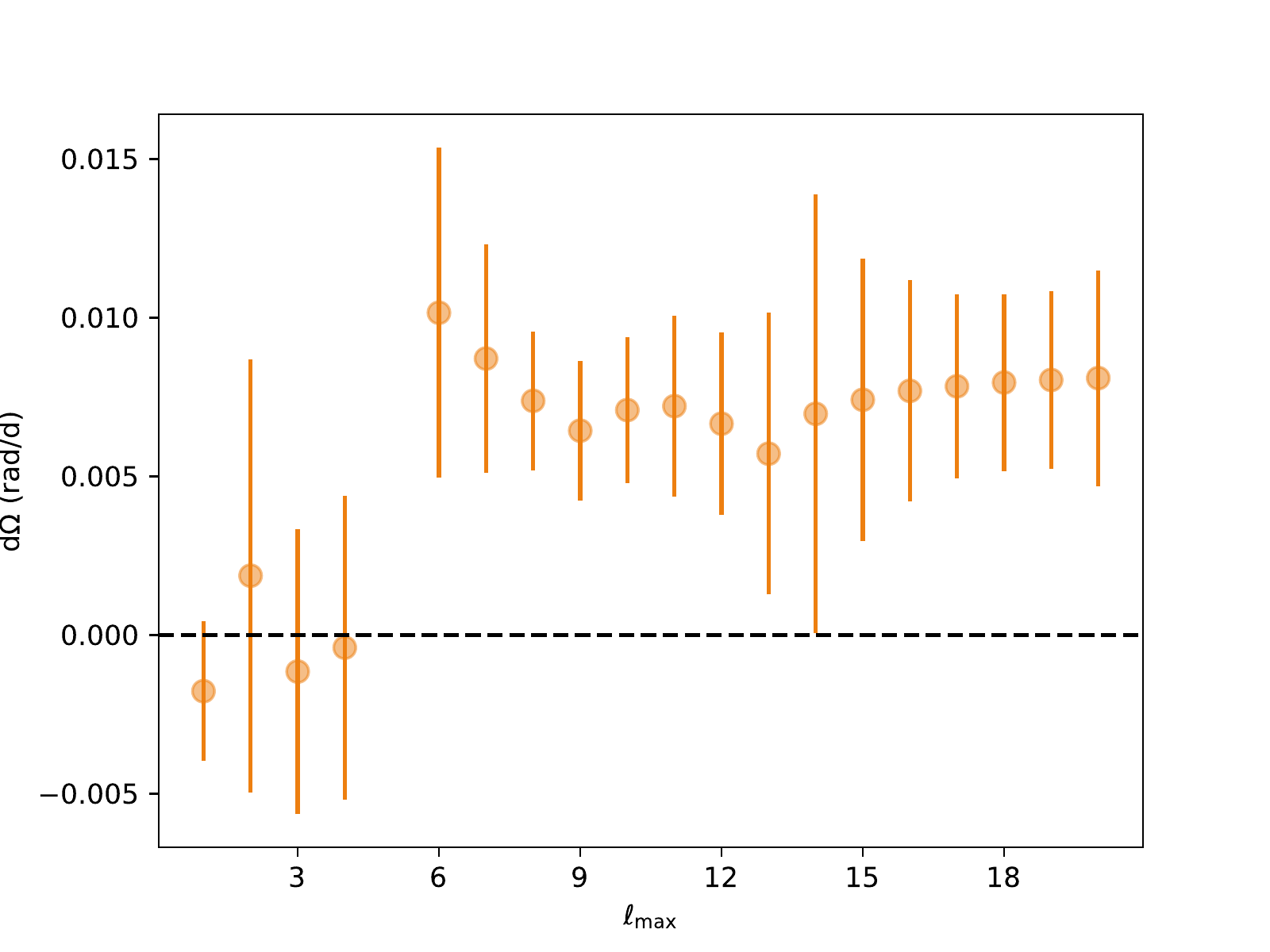}
\caption{Upper panel: values of the differential rotation parameters obtained for different input parameters of the ZDI code. The red symbol shows the reference model. Orange points were calculated with different values of the inclination angle (30$^{\circ}$ and 40$^{\circ}$), and the blue symbol is obtained with a different value of the target entropy (which imposes a slightly larger $\chi^2$, equal to 1.12). Bottom panel: $d\Omega$ values obtained for different values of $\ell_{\rm max}$.}
\label{drotFluctuations}
\end{figure}

We stress that error bars estimated from $\chi^2$ landscapes through the procedure of \cite{press92} tend to be overestimated whenever maximum-entropy inversion is involved, as pointed out by \cite{petit15}. The reason is that the maximum entropy regularization acts as a de-noising procedure, so that error bars estimated through Monte Carlo simulations are smaller than what $\chi^2$ variations may suggest. We provide here the $\chi^2$ error bars anyway, as they are close in magnitude to biases discussed in the two paragraphs below (as previously found by other authors, e.g. \citealt{waite17}). Maximum-entropy estimates of differential rotation parameters are also known to generate systematic measurement offsets, as shown using simulated observations \citep{petit02}. We have estimated the offset for Alhena through simulated observations with characteristics matching our observing situation (phase coverage, rotational broadening, inclination, noise level) and find that $d\Omega$ is likely to be underestimated by about 1\%, while the offset on $P_{\rm eq}$ is negligible.     

The exact position of the minimum in the $\chi^2$ landscape is sensitive to input parameters of the ZDI code. To evaluate the stability of the measured $P_{\rm eq}$ and $d\Omega$, we compute new landscapes and vary one input parameter at a time. Here we choose to concentrate on the inclination angle and target entropy, as these two parameters dominate fluctuations in the outcome. The outcome of this test is illustrated in the upper panel of Fig.~\ref{drotFluctuations}. It is found that the parameter values and error bars change from one model to the next, but differences remain within $1\sigma$ from our reference model, and all results provide us with a clear detection of differential rotation.  

Estimates of differential rotation can also be influenced by the level of details that we allow to be reconstructed in the magnetic map. We show in the bottom panel of Fig.~\ref{drotFluctuations} the $d\Omega$ values recovered for increasing values of $\ell_{\rm max}$. We obtain that ZDI models with $\ell_{\rm max} \le 4$ are all consistent with rigid rotation, while models with $\ell_{\rm max} \ge 6$ favor a non-zero surface shear. We interpret this result as  possible evidence that the largest magnetic structures rotate as a solid body, while  surface features  with  $\ell \ge 6$ (where about 11\% of the total surface magnetic energy is stored) are affected by a systematic shear. Relatively large error bars observed for $\ell_{\rm max}$ = 13,14,15 are likely due to a slight overfitting of the data (i.e. cases where a fraction of the noise pattern in Stokes V profiles is wrongly interpreted as small-scale surface features, therefore diluting the differential rotation signal). Note that the ZDI model computed with $\ell_{\rm max}$ = 5 did not feature any well defined minimum in the $\chi^2$ landscape, hence the missing symbol in the plot.

\section{Discussion}
\label{discus}

\subsection{Magnetic field strength and configuration}  

Among intermediate-mass stars, two categories of surface magnetic fields have been identified so far. Strong magnetic fields (polar strengths above a few hundred gauss) are limited to Ap/Bp stars and seem to concern all objects belonging to this stellar class \citep{auriere07}. Ultra-weak magnetic fields (below the gauss level when averaged over the visible stellar hemisphere) have also been reported more recently. The $\lambda$~Bo\"otis star Vega is the prototype of these weakly magnetic objects \citep{lignieres09}, and small peculiar Zeeman signatures were also reported for a small number of Am stars \citep{petit11,blazere16a}. As far as field strength is concerned the two magnetic regimes are very distinct, to the point that the two magnetic domains seem separated by what was presented as a ``magnetic desert" \citep{lignieres14}.  

The three different methods applied here to LSD Stokes V profiles of Alhena give similar values for the dipolar strength ($\sim$ 30 G). This would place Alhena A in the magnetic desert. The lower limit of the magnetic desert was very roughly estimated using magnetic strengths of two stars only (Vega and Sirius A), so that the actual limit may be above the gauss level. However, it is unlikely that a widespread population of magnetic stars could have remained unnoticed up to now while harboring surface fields of a few tens of gauss, while a number of previous spectropolarimetric surveys were already sensitive enough to uncover such objects \citep{shorlin02,auriere10,makaganiuk11,wade2016}. In this respect, Alhena seems to be atypical among intermediate-mass stars in general, and among Am stars in particular. Its only known sibling identified in the same mass domain is HD\,5550 \citep{alecian16}, an Ap star in a close binary system (while the magnetic desert may be more populated in the massive star regime, e.g. \citealt{blazere15,fossati15}). 

Beside field strength, other magnetic properties of Alhena can be more easily reconciled with known magnetic A stars, on both sides of the magnetic desert. Its prominent dipolar component is reminiscent of the typical large-scale magnetic geometry of Ap stars, and the non-negligible level of complexity of the surface field has also been observed in some Ap stars (e.g. \citealt{kochukhov04}). Below the magnetic desert, complex surface magnetic features were also reported for Vega using ZDI inversion \citep{petit10}.  

All other Am stars studied with the required accuracy to detect ultra-weak magnetic fields displayed peculiar Zeeman signatures \citep{petit10,blazere16a} that may arise from vertical gradients affecting both the magnetic and velocity fields \citep{lopez02}. Alhena\, A is the first Am star where standard polarized signatures are reported (ignoring a limited level of asymmetry discussed in Sect. \ref{sec:diffrot}). As suggested by \cite{blazere16b}, the relatively low micro-turbulence of Alhena may be a clue to understanding this difference. In turn, a lower turbulence may itself be a consequence of its relatively high field strength (bearing in mind that the strong magnetism of Ap stars likely inhibits their surface turbulence \citealt{folsom13}).

\subsection{Evolutionary status}

\subsubsection{Alhena vs other ultra-weak field A stars}

In the literature, Alhena A is classified as a subgiant star whereas Sirius A, $\beta$\,UMa and $\theta$\,Leo are classified as main sequence stars. Therefore, the evolutionary status of Alhena\,A could explain why Alhena exhibits normal signatures contrary to the other Am stars. To test this hypothesis, we determined the evolutionary status of all stars hosting an ultra-weak magnetic field including Vega. We calculated models by interpolating with the SYCLIST tool\footnote{https://www.unige.ch/sciences/astro/evolution/en/database/syclist} in the grid of Geneva stellar evolution models with different metallicities and taking into account the effect of rotation \citep{eggenberger08,georgy13}. To place the Am stars and Vega in the diagram, we used the values of luminosity and temperature found in the literature (see Table~\ref{lum}).

\begin{table}
\caption{Luminosity and temperature with their respective error bars for the Am stars and Vega.}
\centering
\begin{tabular}{c c c}
\hline
Star  & Luminosity (L$_{\odot}$) &  Temperature (K)\\
\hline
\hline
$\beta$ UMa & 63.015 $\pm$ 1.307$^a$ & 9480 $\pm$ 250$^b$ \\
$\theta$ Leo & 141 $\pm$ 1.3$^c$ & 9280 $\pm$ 250$^b$ \\
Sirius A & 24.5 $\pm$ 1.3$^d$  & 9940 $\pm$ 210$^e$\\
Alhena A & 123 $\pm$ 1.3$^f$  & 9150 $\pm$ 310$^g$\\
Vega & 37 $\pm$ 3$^h$ & 9988 $\pm$ 200$^i$ \\
\hline
\multicolumn{2}{l}{$^{a}$ \cite{boyajian12}}&$^{b}$\cite{zorec12}\\
\multicolumn{2}{l}{$^{c}$ \cite{wyatt07} }& $^{d}$ \cite{liebert05}\\
\multicolumn{2}{l}{$^{e}$ \cite{adelman04}} & $^{f}$ \cite{malagnini90}\\
\multicolumn{2}{l}{$^g$ \cite{adelman15}} & $^h$ \cite{yoon10}\\
\multicolumn{2}{l}{$^i$ \cite{yoon08}} & \\
\end{tabular}
\label{lum}
\end{table}

We find that Alhena\,A and $\theta$\,Leo are close to the end of the main sequence. Sirius\,A, Vega, and $\beta$\,UMa are on the main sequence. As a consequence, the shape of the signatures (normal or peculiar) does not seem to depend on the evolutionary status of the stars (see Fig.~\ref{evolution}).

\begin{figure}
\centering
\includegraphics[width=8.5cm, trim= 1.5cm 1.5cm 3cm 3cm, clip]{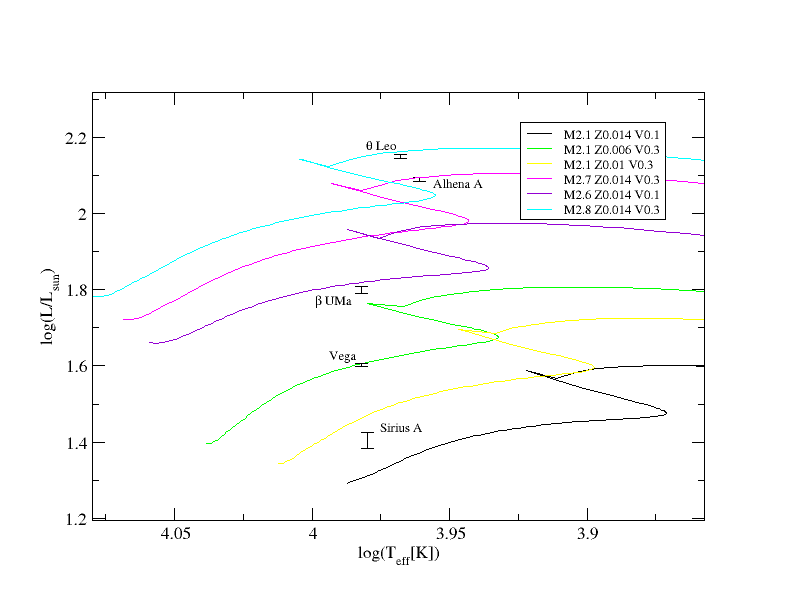}
\caption{HR diagram showing evolutionary tracks computed with the Geneva stellar evolution code, compared to the observed position of the magnetic A stars.}
\label{evolution}
\end{figure}

\subsubsection{Am vs evolved Ap star}

Since Alhena\,A appears at the end of the main sequence, could its magnetic peculiarities  come from the fact that it is an evolved Ap star rather than an Am star? Through the main sequence lifetime, the magnetic field strength at the surface of Ap stars decreases through two processes. First, magnetic flux conservation leads to a weaker field at the stellar surface as the radius increases with time. However, this process alone cannot explain the observed decrease of their field strength with age, so that an intrinsic magnetic decay of unknown origin should simultaneously occur \citep{landstreet08,sikora19}. The 300~G polar field lower limit observed for main sequence Ap stars may thus decrease as well with time. Assuming a doubling in radius during the main sequence, the Terminal Age Main Sequence (TAMS) minimal field could drop below 100~G, which could possibly account for the magnetic strength observed on Alhena, although this would force Alhena to initially belong to the most weakly magnetic fraction of Ap stars. Moreover,  chemical peculiarities decrease over the main sequence life of Ap stars \citep{bailey14}, which agrees with the weak chemical peculiarities of Alhena. 

We also observed variations of the equivalent width of LSD Stokes I profiles. Such a behavior is commonly observed in Ap/Bp stars \citep{landstreet17}. The temporal modifications witnessed in Stokes I are likely due to photospheric spots (chemical and/or temperature spots). If this suggests a similarity between Alhena and Ap stars, such surface structures are also not completely unexpected in Am stars, as rotational modulation was previously reported from Kepler light curves of Am stars by \cite{balona13}, suggesting the widespread presence of low contrast spots in this class of stars. The detectable spectroscopic variability reported here, above what is typical of Am stars, may simply be related to the abnormally high field strength of Alhena (for an Am star).  

\subsection{Differential rotation}

The ZDI model obtained for Alhena offers a significantly better fit to the data when latitudinal differential rotation is included in the inversion procedure. The detection of differentially rotating surfaces is common in cool active stars \citep{barnes05, ammler12}.  Similar searches have been so far unsuccessful when conducted on more massive, strongly magnetic stars \citep{donati06, briquet16}, while intermediate mass stars without strong surface magnetism seem to be differentially rotating \citep{balona16}. This lack of positive results is also consistent with the long-term stability of magnetic geometries of Ap stars (see e.g. \citealt{silvester14} for $\alpha^2$~CVn), which is interpreted as a capacity of their magnetic field to inhibit large-scale flows \citep{mathis05,zahn11,auriere07}.

In this context, finding evidence for a latitudinal shear at the surface of Alhena came as a surprise. Although small compared to the solar shear, it is sufficiently strong to be consistently recovered when varying a number of input parameters of the ZDI code. The test illustrated in the bottom panel of Fig. \ref{drotFluctuations} highlights that the largest-scale magnetic structures in the magnetic map (up to $\ell$ = 4) rotate as a solid body, while the shear signal is confined to surface structures sufficiently small to contribute to spherical harmonics modes above $\ell$ = 6.

With a dipole strength of about 30~G, the large-scale field is above the critical field strength \citep{zahn11,spruit99} necessary to stop the differential rotation in the radiative envelope over a timescale shorter than the age of Alhena. Therefore the differential rotation must be confined to a thinner region close to the surface. To test whether this differential rotation is produced by convection in a thin sub-surface layer, we calculated a stellar evolution model of Alhena\,A and computed the convective Rossby number.

The stellar model was computed with the CESTAM evolution code \citep{lebreton08,marques13}. We used the OPAL opacity tables \citep{iglesias96} completed at low temperatures by the Wichita opacity data \citep{ferguson05}, the OPAL2005 equation of states \citep{rogers02}, and nuclear reactions from the NACRE compilation \citep{angulo99} except for the $^{14}$N(p,$\gamma$)$^{15}$O reaction, for which we used the LUNA reaction rate given in \cite{imbriani04}. The convection was treated with the CGM formalism \citep{canuto96} with a mixing-length parameter $\alpha_{CGM}$=$0.68$ obtained from a solar calibration. We did not include diffusion in the model computation. We used a solar mixture following the \cite{asplund09} one with $Y_0$=$0.2578$ and $(Z/X)_0$=$0.0195$. The model has a mass of 2.8~M$_\odot$, an effective temperature of 9212~K, and a surface gravity of 3.59. For this model, the surface convective zone has a size of $\Delta R$=4.5 10$^4$ km.

Following \cite{Brunetal2015, Brunetal2017}, it is possible to predict the shape of the latitudinal differential rotation in convective shells as a function of the convective Rossby number. This dimensionless number characterises the relative strength of the inertia of turbulent convective flows and of the Coriolis acceleration. In the case of the superficial convective envelope, we compute it as
\begin{equation}
R_{\rm o}^{\rm c}=\frac{U_{\rm c}}{2\Omega \Delta R},
\end{equation}
where $U_{\rm c}$ is a characteristic convective velocity, $\Omega=2\pi/P_{\rm rot}$ with $P_{\rm rot}=8.975\,{\rm days}$ is the global mean rotation rate of the star at its surface, and $\Delta R$=$4.5 \times 10^4$ km is the thickness of the convection zone, which has been computed above with the stellar evolution model. Following \cite{Brunetal2015,Brunetal2017}, we assume that the convective velocity can be expressed as
\begin{equation}
U_{\rm c}=\left(\frac{L}{{\overline\rho}_{\rm  CZ}R^2}\right)^{1/3},
\end{equation}
where $L$=$123 L_{\odot}$ is the luminosity of Alhena expressed in solar luminosity, ${\overline\rho}_{\rm  CZ}$=9.9 g.cm$^{-3}$ is the mean density in the convection zone, and  $R$=$4.03 R_{\odot}$ is the radius of the star expressed in solar radii. This leads to a convective velocity of the order of $\sim$10$^6$ cm.s$^{-1}$ in good agreement with the convective velocity computed by the stellar evolution code using the mixing-length theory. This leads to a high value of the convective Rossby number, $R_{\rm o}^{\rm c}\approx 114$.

In this framework, \cite{Brunetal2017} predicted that convective envelopes with a low convective Rossby number ($R_{\rm o}^{\rm c}<0.1$) are developing band-like cylindrical differential rotation as in giant planets. Convection zones with a moderate convective Rossby number ($0.1<R_{\rm o}^{\rm c}<1$) are hosting solar-like conical differential rotation with an equatorial acceleration. Finally, {the same authors suggest that} convective envelopes with high values of $R_{\rm o}^{\rm c}>1$, like Alhena, are hosting potentially anti-solar differential rotation with a polar acceleration, to the contrary of Alhena. Moreover, \cite{Gastineetal2014} demonstrated that for high values of $R_{\rm o}^{\rm c}>10$, as in the case of Alhena, this anti-solar differential rotation should become very weak. As a consequence, the observed differential rotation with an equatorial acceleration cannot be attributed to the dynamics of the sub-surface convective envelope.

In addition, tidally-induced angular momentum exchanges are negligible in the system of Alhena as explained at the end of Sect.~3. As a consequence, and assuming that the measured shear is reliable, the observed solar-like differential rotation should be explained by another, as yet unidentified, physical mechanism.

\section{Conclusions}\label{conclu}

We confirmed that Alhena\,A is magnetic and we determined its surface magnetic properties thanks to different methods. An inclined dipole model was used to reproduce the Stokes V line profiles, highlighting a polar field strength of $\sim$ 30 G. The magnetic field of Alhena\,A is weak, however it is stronger than the ultra-weak fields discovered on Vega and on other Am stars, and locates the star in the magnetic desert \citep{auriere07}. However, the limits of the magnetic desert are currently not well defined. Whether Alhena\,A is indeed an Am star or could possibly be a weakly-magnetic evolved Ap star also remains to be investigated further.

In addition, the inclination and the obliquity angles of Alhena are low, which explains why the Stokes V profiles change only slightly over the course of the observations. Nevertheless, a rotational period of 8.975 days was identified using intensity line profile variations. 

The ZDI model unveils small scale magnetic structures and  suggests that the surface magnetic field is sheared by differential rotation, with a difference in rotation rate between high and low latitudes about 85\% weaker than solar. This solar-like differential rotation is not expected considering the strength of the detected magnetic field which should freeze differential rotation in the envelope. Moreover, it is not consistent with the convective Rossby number found for Alhena, since high values of the Rossby number should correspond to a weak anti-solar differential rotation regime. The physical process that produced the observed differential rotation is thus not yet identified.  The consequence of this shear on the longer-term stability of the field is also unclear, and future observations may  help us firmly confirm the reality of surface differential rotation and tell us if some long-term variability of the field geometry is experienced by Alhena.

Alhena\,A remains an intriguing and key object in the study of weak magnetic fields. Accumulating more observations is needed to better understand its variations and peculiar properties.

\section*{Acknowledgements}

We thank Richard Monier for useful discussions. The authors acknowledge support from the ANR (Agence Nationale de la Recherche) project Imagine and from PNPS (Programme National de Physique Stellaire). This research has made use of the SIMBAD database operated at CDS, Strasbourg (France), and of NASA's Astrophysics Data System (ADS).  We are grateful to our referee, Evelyne Alecian, for a number of incisive comments that helped to clarify and strengthen our results.

%%%%%%%%%%%%%%%%%%%% REFERENCES %%%%%%%%%%%%%%%%%%

%\bibliographystyle{mnras}
%\bibliography{alhena} 

%%%%%%%%%%%%%%%%%%%%%%%%%%%%%%%%%%%%%%%%%%%%%%%%%%
\appendix
\section{}

\begin{figure*}
\centering
\includegraphics[scale=0.6,trim=0cm 1.5cm 0cm 3cm, clip]{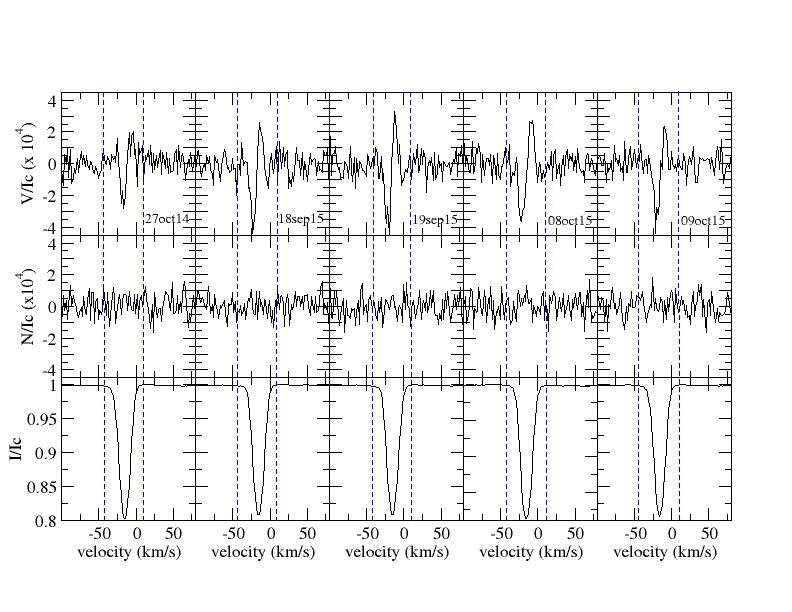}
\includegraphics[scale=0.6,trim=0cm 1.5cm 0cm 3cm, clip]{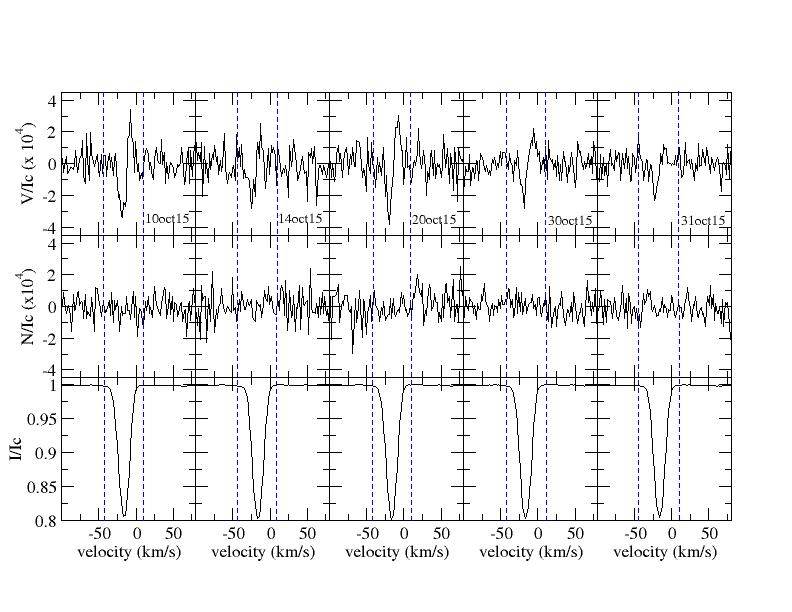}
\caption{LSD Stokes I profiles (bottom), Stokes V (top), and null N (middle) profiles of Alhena.}
\ContinuedFloat
\label{lsd_annexe}
\end{figure*}

\begin{figure*}

\centering
\includegraphics[scale=0.65,trim=0cm 1.5cm 0cm 3cm, clip]{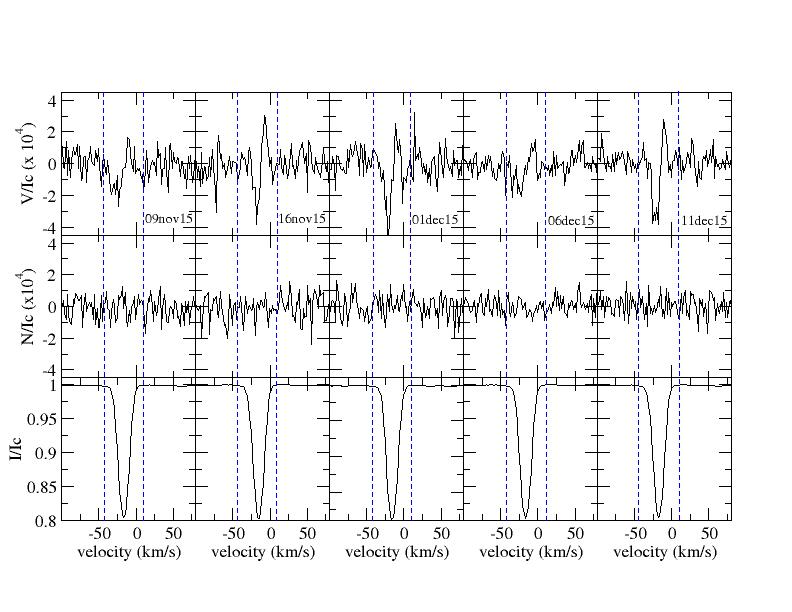}
\includegraphics[scale=0.65,trim=0cm 1.5cm 0cm 3cm, clip]{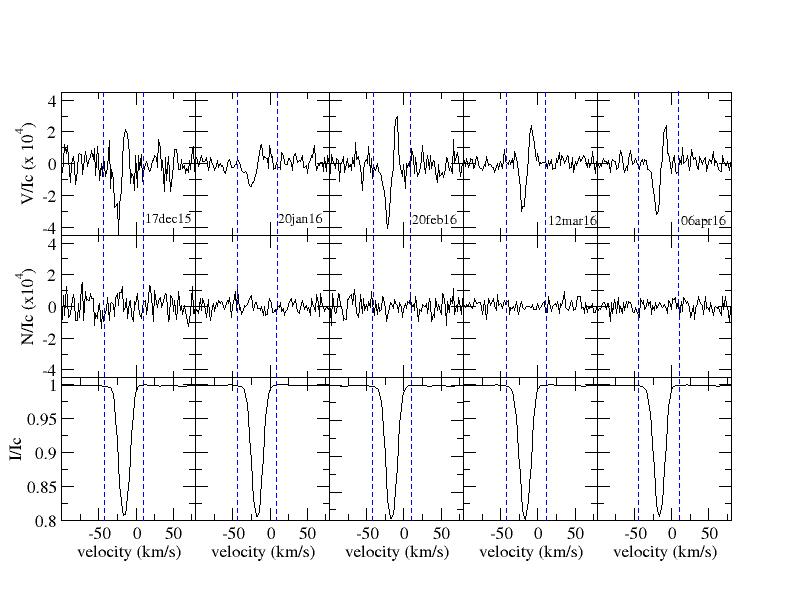}
\caption{Continued}
\ContinuedFloat
\end{figure*}

\begin{figure*}
\centering
\includegraphics[scale=0.65,trim=0cm 1.5cm 0cm 3cm, clip]{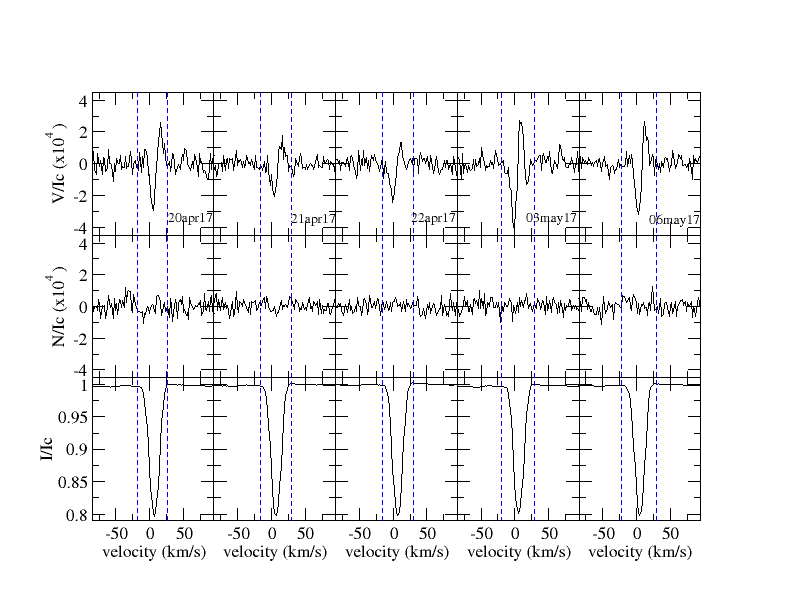}
\caption{Continued}
\end{figure*}

\begin{figure}
\includegraphics[scale=0.5]{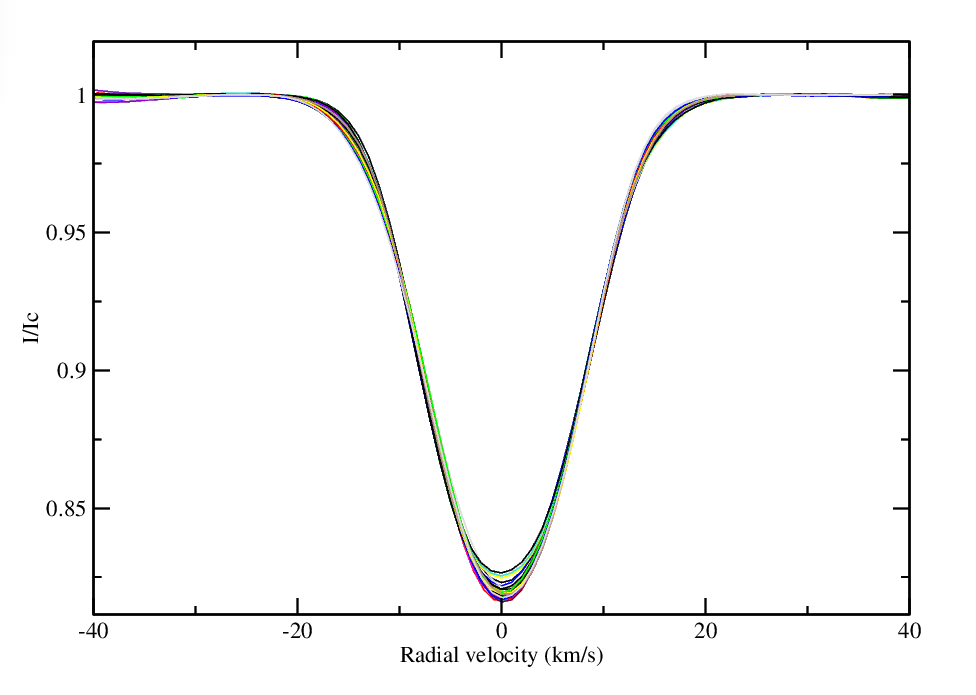}
\caption{LSD Stokes I profiles for all observations}
\label{LSDI}
\end{figure}

% Don't change these lines
\bsp	% typesetting comment
\label{lastpage}
\end{document}